\documentclass[10pt,journal,compsoc]{IEEEtran}

\usepackage{graphicx}

\graphicspath{{figures/}{pictures/}{images/}{./}} 

\usepackage{microtype}                 
\PassOptionsToPackage{warn}{textcomp}  
\usepackage{textcomp}                  
\usepackage{mathptmx}                  
\usepackage{times}                     
\usepackage{cite}                      
\usepackage{tabu}                      
\usepackage{booktabs}                  
\usepackage{multirow}
\usepackage[ruled]{algorithm2e} 

\usepackage{tabularx}
\usepackage{amsmath}
\usepackage{subcaption}
\usepackage{xcolor}

\newcommand{\zfl}[1]{\textcolor{black}{{#1}}}

\newcommand{\neil}[1]{\textcolor{black}{{#1}}}
\newcommand{\nd}[1]{\textcolor{black}{{#1}}}
\newcommand{\hk}[1]{\textcolor{black}{{#1}}}
\newcommand{\hkminor}[1]{\textcolor{black}{{#1}}}
\newcommand{\degree}[1]{°\ }
\DeclareUnicodeCharacter{2218}{\degree}

\author{Roy G. Biv\thanks{e-mail: roy.g.biv@aol.com}\\ %
        \scriptsize Starbucks Research %
\and Ed Grimley\thanks{e-mail: ed.grimley@aol.com}\\ %
     \scriptsize Grimley Widgets, Inc. %
\and Martha Stewart\thanks{e-mail: martha.stewart@marthastewart.com}\\ %
     \parbox{1.4in}{\scriptsize \centering Martha Stewart Enterprises \\ Microsoft Research}}

\begin{document}



\title{360\degree\  Stereo Image Composition \\ with Depth Adaption}

\author{Kun~Huang,
        Fang-Lue~Zhang,~
        Junhong~Zhao, 
        Yiheng~Li, 
        and~Neil~Dodgson
\IEEEcompsocitemizethanks{\IEEEcompsocthanksitem K. Huang, F.-L. Zhang, Y. Li and N. Dodgson are from the School of Engineering and Computer Science, Victoria University of Wellington, New Zealand. \protect\\
E-mail: \{kun.huang, fanglue.zhang, yiheng.li, neil.dodgson\}@vuw.ac.nz
\IEEEcompsocthanksitem Junhong Zhao is with CMIC, Victoria University of Wellington
\IEEEcompsocthanksitem Fang-Lue Zhang is the corresponding author.\protect\\}
\thanks{Manuscript received 2022; revised 2023.}}

\markboth{IEEE Transactions on Visualization and Computer Graphics }%
{Huang \MakeLowercase{\textit{et al.}}: 360\degree\  stereoscopic image composition}

\IEEEtitleabstractindextext{%
\begin{abstract}
360\degree\ images and videos have become an economic and popular way to provide VR experiences using real-world content. However, the manipulation of the stereo panoramic content remains less explored. In this paper, we focus on the 360\degree\ image composition problem, and develop a solution that can take an object from a stereo image pair and insert it at a given 3D position in a target stereo panorama, with well-preserved geometry information. Our method uses recovered 3D point clouds to guide the composited image generation. More specifically, we observe that using only a one-off operation to insert objects into equirectangular images will never produce satisfactory depth perception and generate ghost artifacts when users are watching the result from different view directions. Therefore, we propose a novel per-view projection method that segments the object in 3D spherical space with the stereo camera pair facing in that direction. A deep depth densification network is proposed to generate depth guidance for the stereo image generation of each view segment according to the desired position and pose of the inserted object. We finally combine the synthesized view segments and blend the objects into the target  stereo 360\degree\  scene. 
A user study demonstrates that our method can provide good depth perception and removes ghost artifacts. The per-view solution is a potential paradigm for other content manipulation methods for 360\degree\  images and videos.  

\end{abstract}

\begin{IEEEkeywords}
Stereoscopic Panoramic Image, Image Composition, Image Synthesis, Virtual Reality
\end{IEEEkeywords}}

\maketitle

\IEEEraisesectionheading{\section{Introduction}\label{sec:introduction}}

\IEEEPARstart{A}{dvances} in virtual reality (VR) and digital media technology have allowed people to virtually teleport to a virtual environment. This immersive experience provides tremendous opportunities in entertainment, education, and enriched experiences not directly accessible owing to safety or cost \cite{laga2020survey}. An economical way to construct such a virtual scene is to capture omnidirectional stereo images or videos from the real world. Therefore, there have been emerging research interests in 360\degree\ image and video processing for better immersive experiences in VR applications. But the question of how to manipulate the content of 360\degree\ stereo images remains less investigated. 
As a fundamental task in content manipulation, seamless image composition and cloning have been well-studied in the computer graphics and vision communities, especially for 2D images and videos \cite{Jue_2007,farbman2009coordinates,luo2012perspective,dai2021learning}. However, as demonstrated in the most recent 360\degree\ image/video processing work, such as stabilization \cite{tang2019joint}, depth estimation \cite{wang2020360sd}, optical flow estimation \cite{li2022deep}, and edit propagation \cite{zhang2021efficient,zhang2022fast}, the methods designed for normal 2D images cannot be easily extended to work for 360\degree\ images, because of their incorrect spatial relationship measurement in the spherical domain. 
\begin{figure}[b!]
 \includegraphics[width =0.9\linewidth]{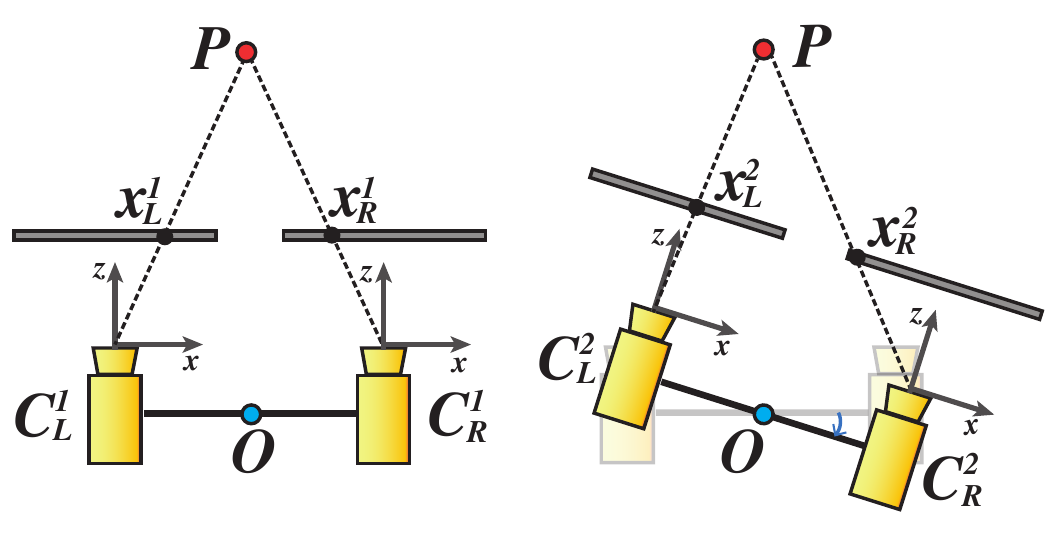}
 \caption{The camera positions when watching stereo panoramas in a VR headset. $P$ is a scene point, $O$ is the virtual position of the user. When the user rotates their head (right), the two cameras are actually rotated about $O$, not their own centers, making the disparity $(x_L^1-x_R^1)$ different from $(x_L^2 - x_R^2)$. This means that an image captured with this camera model has depth errors and ghosting when viewed from any rotated position.}
 \label{fig:exp}
\end{figure}

Besides the typical problems that any image composition method has to cope with, such as gradient mismatch and complex object boundaries, there is an additional challenge with 360\degree\ stereo images: the consistency of the depth perception when the user is focusing on any part of the composited result. That issue can be neglected in planar stereo image composition \cite{tong2012stereopasting}
where a pair of camera positions are defined to look at the scene center, since the field-of-view (FoV) is limited in 2D images. 
However, 360\degree\ stereo images allow users to rotate their view directions to focus on an arbitrary region of the scene. The 360\degree\ images/videos are pre-loaded for the left and right eyes and played by directly projecting the left/right panorama to the left/right viewport for efficiency. If the stereo composition is only conducted as a one-off operation for a predefined user position, i.e., directly pasting the source regions from the left-view and right-view to the stereo equirectangular images for the final result, the perceived depths will not be correct unless the user's view direction is the same as the predefined cameras. Fig. \ref{fig:exp} demonstrates the issue: When a user rotates their head with a VR headset, the depths of scene points vary, so the fixed disparity of a stereo pixel pair generated by the one-off composition can never satisfy all possible view directions. 

In this paper, we propose a novel method to insert stereo objects into a target 360\degree\ stereo image with a convincing appearance and well-preserved depth information when viewers change their viewing orientation. Our method uses the estimated depth information of both source and target stereo images to guide the generation of the stereo pair of inserted objects, ensuring the correct geometry when watching the object inserted at an arbitrary 3D position. For addressing the aforementioned depth inconsistency issue, we propose a solution where the generated disparities of stereo pixel pairs can fit different view directions of the virtual view pair. Instead of using a single pair of camera positions when generating the left and right panoramas, our camera model uses multiple pairs of camera positions facing in different directions. 
For different parts of the object, we separately generate the stereo content using the pairs of cameras looking in each direction and then combine the content of all the parts in the final left/right panoramas. More particularly, we build a deep neural network to learn to generate dense depth maps and object masks to produce high-quality stereo content when the object pose changes in the composited results, outperforming all the previous stereo image composition methods. In our user study, we find that our method offers the best depth perception, especially when the inserted object covers a large FoV in the final result.

Our contributions are as follows:
\begin{itemize} 
\item An omnidirectional stereo image composition algorithm, which can composite a stereo object into a 360\degree\ panoramic background for VR applications. Our method has good fidelity, ensuring the fundamental 3D geometries of the inserted objects by guiding the content manipulation in 3D space. 

\item A novel solution to address the depth perception issue in stereo panorama content generation. We use a camera model that is more suitable to the geometry of stereo panoramas than a model that assumes projection onto a single plane. 

\item A deep model that is able to synthesize dense and accurate depth maps and object masks to facilitate stereo image generation for different object poses.

\end{itemize}



\section{Related Work}

Our work involves efficient compositing of 3D objects into stereoscopic 360\degree\ panoramas. We briefly cover the key related work in image manipulation, 360\degree\  image processing, stereoscopic editing, and 3D object manipulation.

\begin{figure*}[ht!]
 \includegraphics[width = \textwidth]{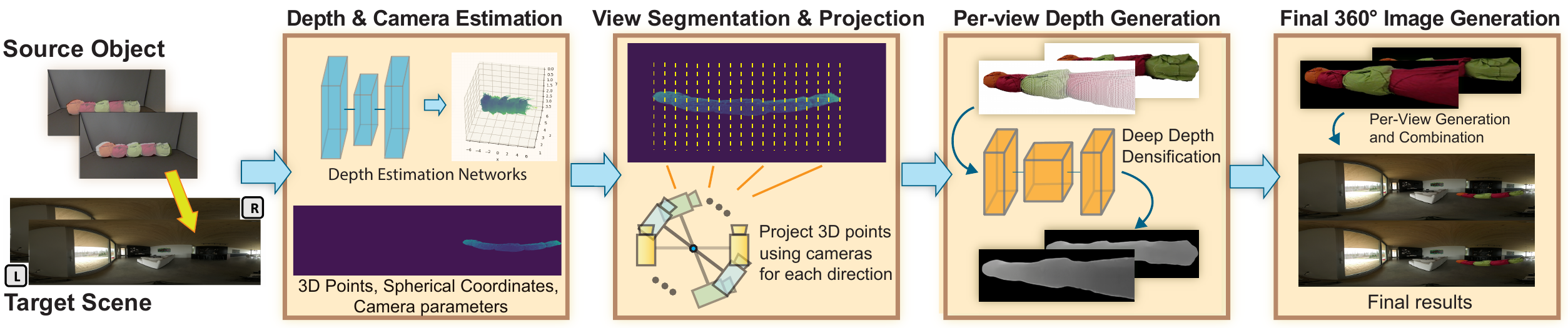}
 \caption{The pipeline of our method. Given a stereo object, our approach manipulates the image content with guidance from 3D space to avoid distance metric issues when compositing the object into the target omnidirectional stereo background images. }
 \label{fig:f3_1}
\end{figure*}

\subsection{Image Matting, Composition, and Segmentation}
Image composition is a basic operation for content manipulation, used initially for film and video production \cite{fielding1965technique}. Early methods focused on providing intelligent scissors for object segmentation to composite \cite{mortensen1995intelligent, mortensen1998interactive}. In recent decades, alpha matting \cite{Jue_2007} and gradient-domain methods \cite{perez2003poisson} have become mainstream approaches for composition. Matting allows us to extract accurate boundaries with transparency values of foreground pixels for realistic object insertion \cite{smith1996blue,chuang2001bayesian}. Gradient-domain methods, such as Poisson Blending \cite{perez2003poisson}, help find a smooth transition between the background and the inserted foreground. Previous work also focused on various aspects of image blending, such as the environment lighting effects \cite{zongker1999environment, wenger2005performance, unger2003capturing}. 
More recently, deep learning-based approaches have been proposed to increase the accuracy of the extracted soft masks of alpha matting \cite{xu2017deep,dai2021learning, ding2022deepmatting} or improve the visual consistency between composited foreground and the target background \cite{cong2020dovenet,wu2020jmnet,li2022bridgingmatting}. The above methods handle 2D planar images very well. But they are not able to generate satisfactory results with stereo 360\degree\ images since an appropriate depth perception cannot be guaranteed. 

\subsection{360\degree\  Image Analysis and Processing}
A great deal of recent work has attempted to understand and process 360\degree\ images and videos for better immersive experiences in VR applications. To provide better 3D information for mixed reality applications based on 360\degree\  videos, Feng et al. \cite{feng2022depth360,feng2020foreground} and Wang et al. \cite{wang2020360sd} proposed deep depth estimation networks working on the spherical domain and built large panorama datasets for training their models. Deep learning techniques have also been used effectively for the semantic understanding of 360\degree\  images, including saliency detection \cite{360Saliency}, object recognition \cite{360recognition}, and indoor holistic scene understanding \cite{sun2021hohonet}. Li et al. \cite{li2021lighting360} developed a method of lighting and geometry estimation from 360 panoramic stereos. In the work of Li et al. \cite{li2022deep}, the dense correspondence estimation for 360\degree\  videos is improved by fusing the information of different sphere-to-plane projections. Although these techniques are capable of processing spherical 360\degree\  images properly, they are not able to be directly applied in stereo 360\degree\  image generation. Some researchers focus on omnidirectional view-synthesis from 360\degree\ image sequences to provide 6-DOF immersive experiences by explicitly \cite{Chen20226dof360} or implicitly \cite{barron2022mip360} reconstructing 3D geometry. Zhao et al. \cite{zhao2021adaptive} and Xu et al. \cite{xu2022rendering} proposed to use convolutional neural networks to predict 360\degree\  HDR images for a better illumination effect when inserting virtual objects into a target scene. But they are not designed for manipulating stereo image content. To improve the interactive experiences, researchers presented methods for allowing a better user simulation \cite{martin2022scangan360} and adding social features to the VR video player \cite{li2022bullet360, 9417718}. We focus on providing richer experiences by allowing the user to modify scene content.  

\subsection{Stereoscopic Image Editing}
Stereoscopic image editing has attracted much research in the past decade, initially prompted by the needs of stereoscopic 3D film production \cite{mendiburu20123d}. Wang et al. \cite{wang2011stereobrush} investigated a novel workflow called \textit{Stereo\-Brush} for users to convert a 2D stereoscopic image to 3D instantly by drawing strokes on the 2D image. Other research focuses on stereoscopic editing for stereo visual comfort by applying \hkminor{the mesh-based} image warping manipulation methods to adjust the image structure. Tong et al. \cite{tong2012stereopasting} proposed a novel system named \textit{Stereo\-Pasting}, inspired by \textit{Stereo\-Brush}. It solves the stereoscopic composition task using an energy minimization warping formula. Users get instant feedback while painting strokes on the 2D foregrounds. Luo et al. \cite{luo2012perspective} developed an algorithm for seamless stereoscopic image cloning, which manipulates on both color appearance and perceived depth. It estimates the disparity in the gradient domain to make the disparities of the cloning region continuous at the boundary, and also adjusts the shape and size of the cloning area by applying a perspective-aware warping \hkminor{with the constructed mesh based on the estimated disparity.}
\hk{Du et al. \cite{du2013changing} introduced \nd{a 2D warping method for adjusting stereoscopic imagery}, enabling users to perceive stereopsis in a new view. \nd{They use} feature correspondences and straight-line constraints to guide the warping process, \nd{treating it} as a quadratic energy minimization problem. However, \nd{these previous methods (\textit{Stereo\-Pasting}, Luo et al., Du et al.)} are not plausible to deal with large perspective differences or occlusion between foreground and background because the generated disparities between corresponding pixel pairs can be significantly distorted. Furthermore, \hkminor{methods relying on mesh-based warping} are unable to generate new \nd{image data} that is required to fit the new perspective, if it was not present in the original input. They are also incapable of hiding information that is occluded under the new perspective \hkminor{because the region that should be occluded can only be narrowed down in the results}.}

Other research in stereoscopic image processing focuses on how to estimate accurate disparity/depth maps \cite{brown2003advances}. Recent advances in stereo depth estimation consist of deploying deep networks embedding all steps of traditional pipelines and combining effective learning modules \cite{Tosi_2018_ECCV,li2020revisiting}. The estimated stereoscopic correspondences are used to conduct view-consistent image enhancement operations via deep networks, such as neural style transfer \cite{chen2018stereoscopic}. Due to the special distortions of 360\degree\ stereo images, deep networks that are delicately designed for estimating depth maps for stereo panoramic images were developed by Wang et al. \cite{wang2020360sd}, which assume vertical parallax between two views. Here, we use the depth information of the target scene and the geometric structure of the foreground object to allow the elements to be composited naturally while keeping a correct sense of occlusion and perspective. 

\subsection{Object Manipulation in 3D Space}
Our work is also related to object modelling and editing methods in 3D for 2D images. Previous work that focused on editing a target object in 3D space needs either to reconstruct its basic geometry structure \cite{chen20133} or to use point clouds \cite{mandikal20183d}. Van der Heuvel \cite{van19983d} and Criminisi et al. \cite{criminisi2000single} introduced techniques of 3D reconstruction from single images, particularly for artificial objects that usually contain substantial prior geometric knowledge. Images of humans, which lack this geometry, do however contain prior structural knowledge that can be used to reconstruct free-form and texture-mapped models \cite{zhang2002single}. The reconstruction and manipulation of human models have been significantly advanced by neural network-based technologies \cite{su2021nerf,athar2023flame}, where the geometry information is implicitly predicted and interpreted. To improve the fidelity of object insertion in a VR environment, Morioka et al. \cite{morioka2016handy} proposed a method to let the inserted 3D object reflect the real-time lighting changes of the scene. 

We choose to use point clouds to model the 3D foreground object because we can obtain depth information from stereo images of the object.
\hk{The point clouds in 3D space enable flexibility when users edit the orientation or scale of the inserted objects without altering the underlying geometry structure.} Furthermore, these 3D points are used to guide the warping and interpolation to generate the composited regions in the target equirectangular stereo pair. However, none of the above methods consider the depth consistency issue when the scene is presented as a 360\degree\ stereo image. In this paper, we propose a paradigm that can produce correct depth perception from an arbitrary view direction in a manipulated 360\degree\ stereo image.



\section{Overview} 
\begin{figure*}[t]
 \centering
 \includegraphics[width =1\textwidth]{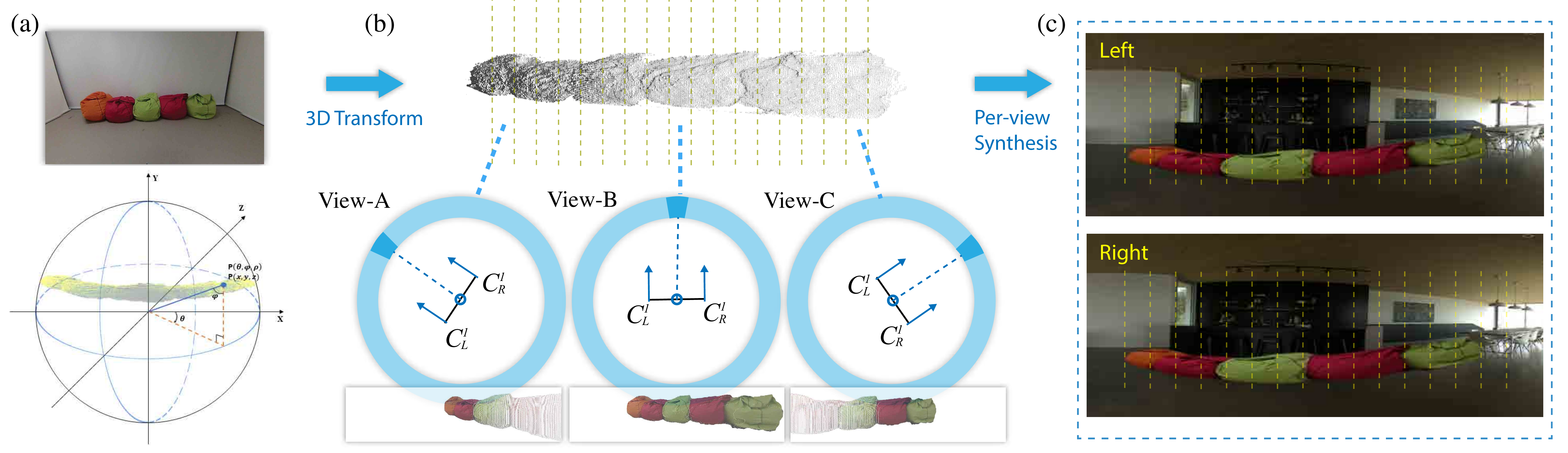}
 \caption{Per-view image generation. (a) Spherical coordinate representation. (b) Per-view projection. (c) The stereo pair is generated with guidance from the densified depth maps for each view segment.}
 \label{fig:view_dependent}
\end{figure*}

Fig. \ref{fig:f3_1} shows the pipeline of our method for compositing a stereo object into a target omnidirectional stereo background image. There are two key points in our method. \hk{First, we transform the common 2D image \nd{to 3D space, using an estimated depth map}. After manipulating the image content with guidance, we further project the 3D coordinates to their spherical positions on the omnidirectional stereo (ODS) images to avoid the inconsistency distance metric issue in different image domains, \nd{as noted by Zhang et al.~\cite{zhang2021efficient}}.}
Second and more importantly, we apply per-view projection from 3D space to ensure appropriate disparities for different parts of the object. The 3D point clouds of the input stereo object and the target scene are reconstructed from the estimated disparities. According to the desired position and size of the inserted object, we transform the 3D points to a spherical coordinate system $(\theta, \phi, \rho)$ with the user's virtual position as its origin, and segment the point clouds into multiple regions based on the horizontal angle $\theta$. To generate proper depth perceptions for an arbitrary view direction, we build separate virtual camera pairs focused on each region, and apply per-view projections to obtain the initial sparse depth maps on the planar image domain. In our experiment, we find that denser segmentation always leads to higher visual quality. Therefore, we normally choose the smallest interval we can achieve to segment the 3D point cloud, which is the viewing angle covered by one column of the target equirectangular image. We then employ a deep depth densification model to estimate the dense depth maps and their alpha maps for all the view directions. The left and right color images are then generated with the guidance of the depth maps. For each view segment, we find the stereo equirectangular pixels within the segment's FoV and overwrite them with the corresponding pixels in the generated planar image pair for that view.
\hk{Our proposed system also offers more possibilities for addressing the occlusion problem that arises when inserting a source object into a background scene, an area that has been less explored in \nd{previous work} on stereo planar image composition with depth information. Properly handling occlusion is crucial for achieving natural depth perception in the final composite, as it eliminates any depth conflicts between the inserted object and the original panorama. However, creating a 3D scene from a panoramic image remains a challenging task due to the fact that depth information can only be obtained from the front view. Furthermore, current methods for 360\degree\ monocular depth estimation are inadequate in producing accurate real-world predictions because of the significant noise present in the input data.}

\section{Algorithm} 
The inputs of the method are a pair of stereo images of the object to be inserted, with a masked region-of-interest (RoI), and a stereo panoramic image pair as the target scene. The output is the composited image pair with correct depth perception of the inserted object for arbitrary view directions. The critical challenge is to find a proper binocular camera model to project inserted objects to target panoramic images while preserving correct depth perception. Another challenge lies in the generation of a complete and smooth depth map especially when the desired inserted 3D position is different from the original source position.
\subsection{Sparse 3D Reconstruction}
We first estimate the depth map of the foreground object and the region around the desired position of the target scene. For inserting the object into the stereo panoramic target scene represented by equirectangular projection, the user is required to specify the position and the size of the object. We project the RoI of the target scene to a planar image with a default FoV of 60\degree. If the source object is in a stereo panorama as well, we use an FoV of no less than 60\degree\ to cover the horizontal angle extent of the object when projecting the object into a planar stereo crop. We apply Li et al's sequence-to-sequence correspondence perspective deep model, named the STereo TRansformer (STTR), to estimate the disparity map from the stereo pairs \cite{li2020revisiting}.  

Using the predicted horizontal disparities between the input rectilinear stereo image pair, we can generate a depth map of the foreground objects based on camera parameters. Assuming the focal length is $f$ and the baseline between the left and right camera is $B$, the depth values $z$ for the pixels can be calculated from its estimated disparity $d$:  $z = (f \times B) / d$.

Given a pixel $(p_x, p_y)$ on the left-view image with a size of $(W, H)$, we assumed a standard camera model located at the origin looking down the $z$-axis. The 3D coordinates of this point in the world coordinate system are obtained by: 
\begin{equation}
    (x, y, z)=\left(\frac{(p_x-W/2)\times z}{f}, \frac{(p_y-H/2)\times z}{f}, z\right),
\end{equation}
In the recovered sparse point cloud from the stereo pixels, we select the center point of the cloud as the object's reference point to place in the target scene and about which to make any transformations, such as scaling and rotation. It  also helps the user to correctly position the object in the target scene. 
Fig.~\ref{fig:view_dependent} shows an example of a recovered 3D point cloud.

\subsection{View Segmentation and Projection}
\label{sec:ViewSeg}
Having determined the transformations needed to meet the user's desires, including the size, orientation and position of the object, we obtain the 3D point cloud for the object that is to be inserted in the target scene. \hk{In order to tackle the challenge of varying depth perceptions, we partition the point cloud into distinct segments based on the respective viewing directions relative to the user’s position.}
As shown in Fig.~\ref{fig:view_dependent}(a), we first transform  world coordinates to a spherical coordinate representation by:
\begin{equation}
\theta=\arctan\left(\frac{x}{z}\right)\ \ 
\phi=\arctan\left(\frac{y}{\sqrt[2]{x^{2}+z^{2}}}\right), \ \ 
\rho=\sqrt[2]{x^{2}+y^{2}+z^{2}}
\end{equation}
Then we segment the point cloud according to the points' $\theta$-values: we split the range of $\theta$ into $N$ intervals, i.e., each interval covers a range of $\frac{2\pi}{N}$, and segment the points of the object based on the horizontal angle intervals in which they fall. In our experiments, we found that some segments might only contain too few points when the user-specified pose is largely different with the original pose, because the point cloud was recovered from the view of the original stereo image. Therefore, we produce a dense depth map for the desired pose and scale of the target object using the \textit{deep depth densification} network proposed later in Sec.~\ref{sec:ddd}, and then perform view segmentation on the point cloud generated using the dense depth map.  

\begin{figure*}[t]
 \includegraphics[width =\linewidth]{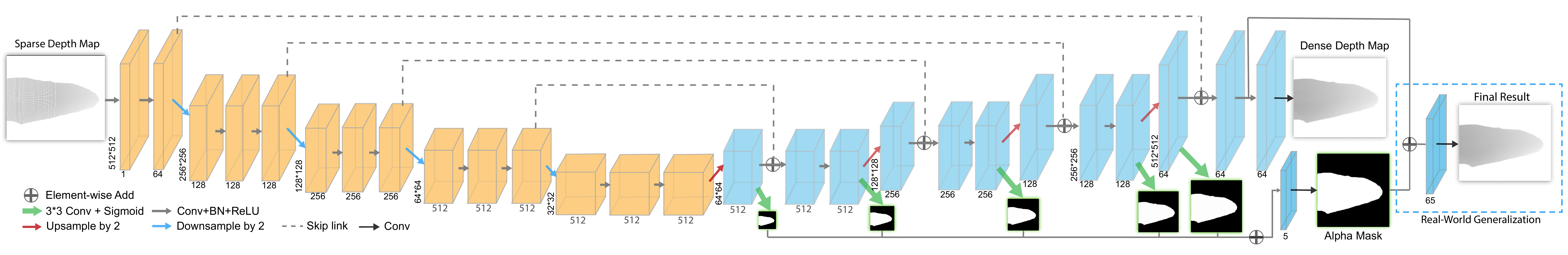}
 \caption{The architecture of the deep depth densification network. We predict a dense depth map for the desired perspective and a soft mask that indicates the pixel's transparency and whether the pixel is solid.}
 \label{fig:network}
\end{figure*}

\

\noindent\textbf{Per-view Projection} 
In this step, we treat each segment of the point cloud with a virtual camera pair that focuses on the individual segment's center. \neil{We generate a series of per-view projected point clouds, each} in their specific camera space (Fig.~\ref{fig:view_dependent}(b)). 
The position and orientation of the user's binocular views are defined as follows. The virtual viewing camera pairs are located on the $xz$-plane of the world coordinate system, with up-vector along the positive $y$-axis. The initial position of \neil{the two cameras}, $C_L^0$ and $C_R^0$ \neil{are located on} opposite sides of the origin on the $x$-axis with a distance $B$ between the two cameras, which are at $C_L^0 = [-B/2, 0, 0]$ and $C_R^0 = [B/2, 0, 0]$, respectively. This initial pair's viewing directions are parallel to the positive $z$-axis. Given $N$ intervals, the system calculates the viewing direction $\theta_{i}$ for the $i$th interval as:
\begin{equation}
    \theta_i= 2\pi i/N-\pi
\end{equation}
Then, the rotation matrix $R_i$ for a such interval is computed based on the viewing direction, which is a rotation about $y$-axis by an angle $\theta_{i}$.
We can define the desired position of $C_L^i$ and $C_R^i$ with associated viewing direction as:
\begin{equation}
    C_L^i = R_{i} C_L^0,\ \
    C_R^i = R_{i} C_R^0
    \label{eqn:camera}
\end{equation}
\hkminor{The projected 3D point cloud in the world coordinate system $p_w$ will be further transferred to each segment space for both views, $Q_L^i$ and $Q_R^i$ with the specific camera matrix are expressed as}:
\begin{equation}
    Q_L^i = [R_{i}| C_L^i] p_w = P_L^i p_w,\ \
    Q_R^i = [R_{i}| C_R^i] p_w = P_R^i p_w
\end{equation}
Using the known intrinsic matrix of both cameras for a specific segment, we then project the point clouds to 2D depth maps. The projected depth maps usually contain holes and gaps when the relative pose of the object to the camera changes. \neil{The following steps (see below) thus estimate} a dense depth map to guide the generation of stereo RGB images for that view direction. Here, instead of just generating a depth map for the corresponding segment, we also generate sparse depths for the neighbouring regions in a certain field-of-view to facilitate the following dense depth map generation for each view.

\subsection{Per-view Depth Generation}
\label{sec:ddd}
To ensure the perceived depths are correct when users are looking at different parts of the composited object, we propose to generate the dense depth maps and corresponding stereo RGB images based on the projected point clouds for each view \neil{direction} separately. The stereo images for each view \neil{direction} will be combined together to generate the final left and right panoramas. 

Given \neil{that the desired pose of an object can vary considerably away from the captured pose}, the projected depth maps \neil{can} be very sparse for \neil{some} view directions of the stereo camera pairs. We \neil{initially} attempted to directly fill the missing 2D pixels on the depth map by morphological interpolation-based methods as described in \cite{ku2018depthCompletion}. However, an interpolation-based method works only when the pose changes are subtle so that the missing points can easily find valid depth values in their neighbourhoods. If the object has a relatively large scaling and rotation, the valid points \neil{are} too sparse to provide sufficient reference values for the missing pixels to interpolate. Moreover, since we need to generate dense depth maps for each view, the long running time of morphological interpolation \neil{has} a large negative impact on efficiency. Thus, we propose a \textit{deep depth densification} network (DDDN) to solve the above issues when generating new depth maps. We show that our network generates dense depth maps efficiently with higher visual quality, especially for objects with sharp geometry features. \neil{To avoid the distortion in the equirectangular representation affecting the learning process}, our deep model works in the 2D rectilinear image domain.

\ 

\noindent\textbf{Architecture}
We build a convolutional generative network that takes a projected sparse depth map as input and predicts dense valid depths and transparencies for the target object. 
As shown in Fig.~\ref{fig:network}, a 2D sparse depth map goes through a U-Net-like architecture to produce a densified depth map. More specifically, we employ the following two learning schemes to improve the quality of generated depth maps: First, we explicitly predict a mask that indicates whether a pixel belongs to the target object in the final image, helping the network to learn whether a pixel should have a valid depth value. Second, we let the decoder learn from multi-scale masks to have a better capability of reconstructing the object's geometry structure. For each scale level, we use a separate convolutional layer, upsampling operation, and sigmoid function to predict the masks at different scales, which are then fused by a concatenation operation at the end with some higher-scale contextual information. \hk{Finally, the output mask and depth map are fused with a concatenation operation, followed by a 1 x 1 convolutional layer and a sigmoid function to generate the final predicted result.}

In our experiment, the above schemes are shown to be effective, especially for improving the depth map quality for those regions with large sparsity. 

\

\noindent\textbf{Dataset} 
\hk{We train and evaluate the above network using both synthetic and real-world datasets. We build a synthetic dataset with ground truth depth maps of rendered objects. }
We use Unity3D to render RGB images and their associated dense depth maps with different positions, scales, and orientations. We select 30 classes of 3D shapes in the ShapeNet dataset \cite{chang2015shapenet}, covering a variety of categories of furniture, vehicle, housewares, and buildings. In each class, we randomly select 50 objects. For each object, we render 20 different poses, producing 20 dense depth maps, $\{D_k\}$, and their corresponding masks, $\{M_k\}$. The original pose for each object is chosen as its reference pose \hkminor{and a point cloud $P_0$ is reconstructed from its depth map $D_0$.} The transformations, between the other poses and the reference pose are then applied on $P_0$ to generate the sparse depth maps, $\{\hat{D}_k\}$. We use the sparse depth map, $\hat{D}_k$, to simulate the input data for the real-world use case, where the sparse depth map can be generated by transforming the reconstructed point clouds from the original stereo object to the desired 3D orientation and position. The corresponding dense map, $D_k$, and mask, $M_k$, can be used to supervise the learning process.

\begin{figure}
\centering
\includegraphics[width =1.00\linewidth]{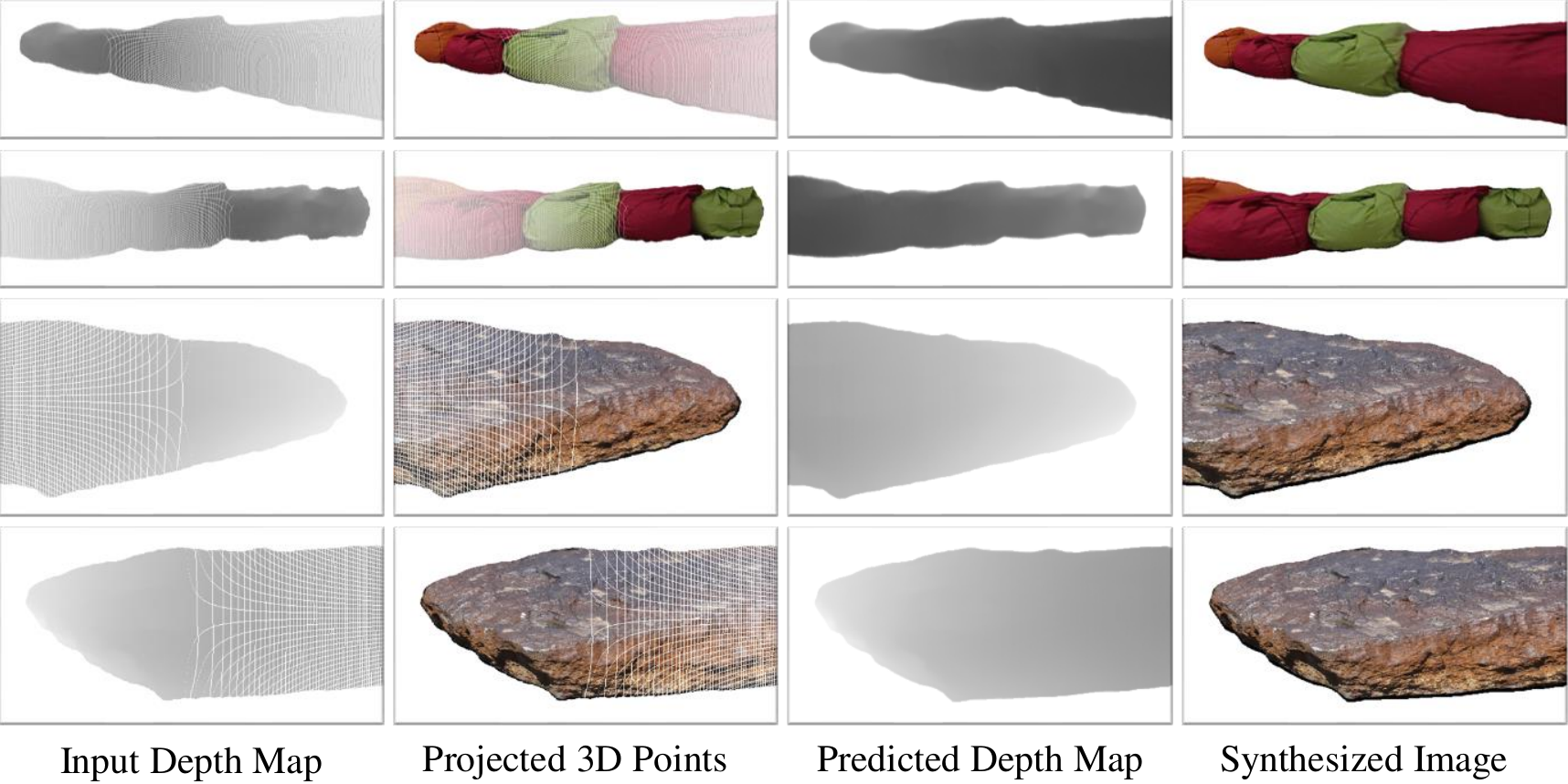}
  \caption{\zfl{The input sparse depth map (first column). The predicted dense depth maps (third column) by our network and synthesized dense RGB images (fourth column) with the guidance of the depth maps. The projected 3D points of the transformed source objects are shown in the second column.} }
  \label{fig:predictRes}
\end{figure}

\hk{\nd{Our real-world dataset is drawn from the HRWSI dataset \cite{Xian_2020_CVPR}. Our} subset covers 15 different labels including living and non-living objects, such as humans, birds, buildings, and chairs, among others. To ensure the quality when generalizing to real-world data, we select 67 main objects that are completely captured within the images. For each of these objects, \hkminor{we construct the point cloud $P_0$ first from its original depth map $D_0$. We then produce} 60 sparse depth maps $\{\hat{D}_k\}$ with corresponding camera pose (rotation $R$ and translation $T$) $\{R^T_k\}$ \hkminor{and original point cloud $P_0$} in 30 different stereo camera-viewed poses. As we lack the ground truth dense depth maps for the transformed sparse depth maps, we adopt a self-supervised approach for generalizing to real-world images. We leverage the corresponding original view of each object as its reference view $D_0$, along with the predicted dense depth maps ${F_d(\hat{D}_k)}$ and the inverse camera pose ${(R^T_k)^{-1}}$ to enable us to carry out the self-supervised process effectively. }

\ 

\noindent\textbf{Training Procedure }
The neural networks are trained to fill the gaps and missing points while preserving the important geometry structures for a sparse depth map. We use the following losses in the training procedure:
\\
(1) The reconstruction loss, $L_r$, an L2 loss applied on the predicted depth map with the ground truth mask, defined as:
\begin{equation}
    L_r = \sum^K_{k = 1}|D_k - F_d(\hat{D}_k) \cdot M_k|_2
\end{equation}
where $F_d(\hat{D}_k)$ represents the predicted dense depth map from the sparse map $\hat{D}_k$ and $M_k$ is the ground truth mask\\
\neil{(2) The mask loss, $L_m$. The network generates masks of different scales, which help it to better learn the global structure of the shape. $L_m$ is the loss of these smaller, subsidiary output masks, which is defined as:}
\begin{equation}
    L_m = \sum^K_{k = 1} \sum^L_{l = 0} |M^l_k - F^l_m(\hat{D}_k)|
\end{equation}
where $M^l_k$ is the ground truth mask for the scale $l$ and $F^l_m(\hat{D}_k)$ is the predicted mask for scale $l$.\\
\neil{(3)} The perceptual loss, $L_p$, applied on the depth map to produce better details. \neil{This makes} the overall loss:
\begin{equation}
    L = \lambda _rL_r + \lambda _mL_m + \lambda _pL_p
\end{equation}
By default, we set $\lambda _r = 0.4$, $\lambda _m = 0.6$, and $\lambda _p = 1.0$ as the weights for different terms. We split our synthetic datasets into a training set of 19950 images and a test set of 9000 images. 
We train our network by 100 epochs or make an early stop when the losses on the validation data stop declining.\\

\hk{For the self-supervised real-world generalization procedure, \hkminor{we first transfer the predicted masked depth map along with the corresponding inverse camera pose back to their depth maps in the original pose. Then, we apply the following losses on the transferred depth map ${F_d(\hat{D}_k)}'$ and its corresponding transferred mask ${F_m(\hat{D}_k)}'$.}}\\
\hkminor{(4) The BerHu loss, $L_b$, applied on ${F_d(\hat{D}_k)}'$ for optimizing depth predictions from the original depth map ${D_0}$ with the corresponding transferred mask: }
\color{black}
\begin{subequations}
\begin{equation}
L_b=
\begin{cases}
|{F_d(\hat{D}_k)}' - {D_0} \cdot {F_m(\hat{D}_k)}'|    
& |{F_d(\hat{D}_k)}' - {D_0} \cdot {F_m(\hat{D}_k)}'|\leq C,\\
\frac{({F_d(\hat{D}_k)}' - {D_0} \cdot {F_m(\hat{D}_k)}')^2 + C^2}{2C} 
& \hfill |{F_d(\hat{D}_k)}' - {D_0} \cdot {F_m(\hat{D}_k)}'| > C,
\end{cases}
\end{equation}
\begin{equation}
C=0.2max(|{F_d(\hat{D}_k)}' - {D_0} \cdot {F_m(\hat{D}_k)}'|)
\end{equation}
\end{subequations}

\hk{(5) The BCEWithLogits loss, $L_{bce}$, applied on the predicted transformed mask with its masked original's mask.
This makes the overall loss: }
\color{black}
\begin{equation}
L = L_b + L_{bce}
\end{equation}
\hk{We split our real-world datasets into a training set of 4020 images and a test set of 1206 images. We do the real-world generalization on our network and apply an early stop when the losses on the validation data stop declining.}

\begin{figure*}[t]
\centering
 \includegraphics[width =1.00\textwidth]{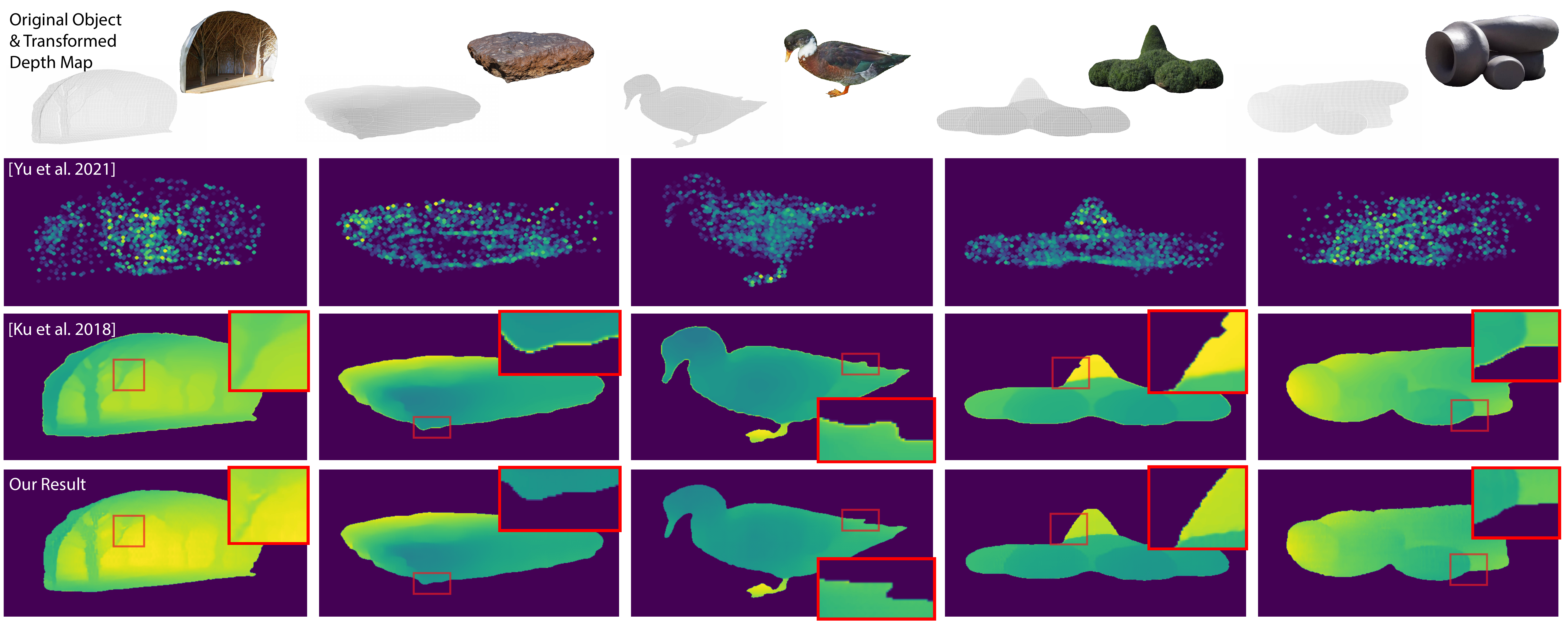}
 \caption{Comparison on depth densification. We show the source objects and their sparse depth maps after 3D transformation in the top row and the generated depth maps by PointTr\cite{yu2021pointr} (Row 2), depth map completion method of \cite{ku2018depthCompletion} (Row 3), and our depth densification network (Row 4).}
  \label{fig:compareDepth}
\end{figure*}

\begin{figure*}[t]
\centering
\includegraphics[width =1.00\linewidth]{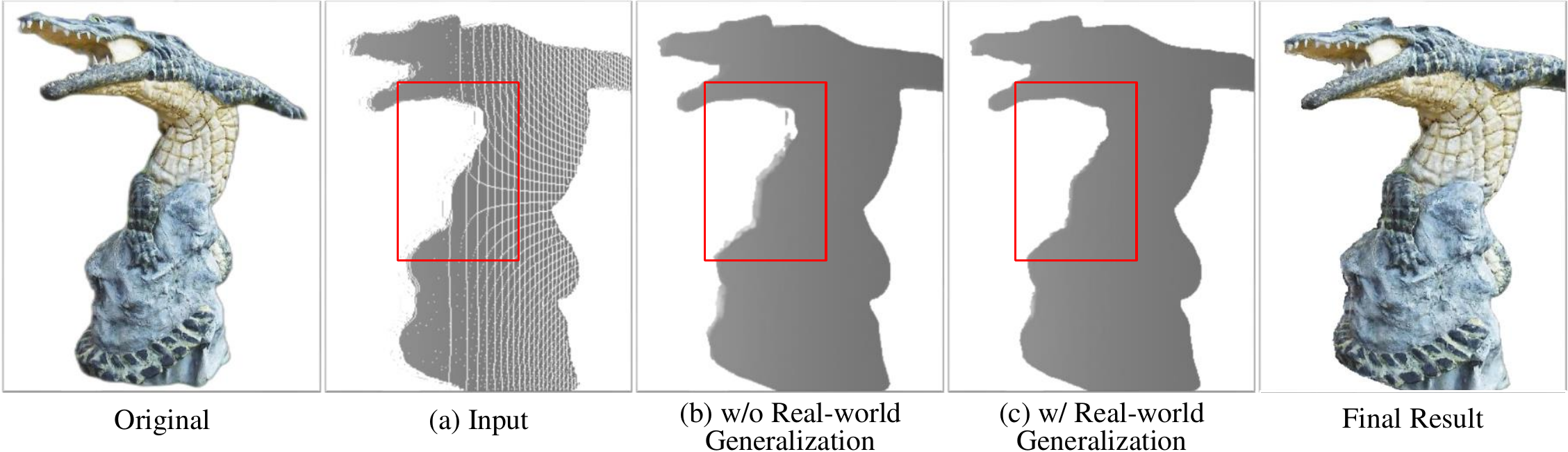}
  \caption{\hk{The input sparse depth map, and the comparison of the predicted results if the real-world generalization is applied. \nd{The red rectangle region shows where real-world generalization makes a significant difference.}}}
  \label{fig:domain_adaptation}
\end{figure*}

\
We show some results of our method in Fig.~\ref{fig:predictRes} and compare our method with some alternatives in Fig.~\ref{fig:compareDepth}. \neil{These demonstrate that it} is hard for interpolation-based methods to properly fill the missing pixels in the boundary parts since the transformed points can be very sparse in the projected depth map and cannot form any continuous line structures to wrap up the object. The use of the mask losses avoids the depth value ``leaking'' to pixels that should be outside of the object's contours and also makes the network more confident when estimating the depth values for pixels with strong geometry features. \hk{In Fig.~\ref{fig:domain_adaptation} and Tab.~\ref{tab:rwgeneralisation}, we compare the performance of our approach with and without the real-world generalization scheme, which has been incorporated to enable better generalization between synthesized and real-world data. This integration has led to a significant reduction in the amount of noise present in the generated depth maps.}

Using such a network, we are able to \neil{rapidly} produce all the dense depth maps where the camera pairs are focused on the centers of different 3D point segments. 

\subsection{Final Stereo Panorama Generation}

Given the dense depth map for each view direction, we re-project each pixel to the original input stereo image to obtain its color value. In the original image, the system applies alpha matting \cite{Germer2020} to generate soft edges for composition. To ensure a seamless composition, we also create an associated alpha mask by copying the alpha value of each referred original pixel. 
For the $N$ segments of the point cloud, we thus obtain $N$ stereo image pairs $\{I^r_n, I^l_n\} (n = 1, 2, ..., N)$ with their alpha masks. To accelerate the process, we \neil{generate the depth map and stereo RGB images only in the} involved part of the object and its neighbouring region, since the other parts, out of the specified view range, will not be used in the final result generation. \neil{We did experiment with} feeding the sparse depth map of the \neil{\emph{entire} object into the neural net, but found that this} does not produce better quality, but rather reduces the resolution of the focused segment of the object in the resultant image. 

\begin{table}
\color{black}
\centering
\normalsize
\begin{tabular}{ccccc}
\hline
\multicolumn{1}{c|}{Method} & \multicolumn{1}{c|}{MAE} & \multicolumn{1}{c|}{RMSE} & \multicolumn{1}{c|}{SSIM} & \multicolumn{1}{c}{PSNR} \\ \hline
\multicolumn{1}{c|}{w/o Gen.} & \multicolumn{1}{c|}{0.9317} & \multicolumn{1}{c|}{5.3009} & \multicolumn{1}{c|}{0.9150} & \multicolumn{1}{c}{35.6401} \\ 
\multicolumn{1}{c|}{w/ Gen.} & \multicolumn{1}{c|}{\textbf{0.7837}} & \multicolumn{1}{c|}{\textbf{4.0211}} & \multicolumn{1}{c|}{\textbf{0.9174}} & \multicolumn{1}{c}{\textbf{39.4472}} \\ \hline
\end{tabular}
\caption{\hk{Quantitative results on real-world datasets with two methods: without and with real-world generalization.}}
\label{tab:rwgeneralisation}
\end{table}

We compose the synthesized left and right views of each segment to the left and right views of the target panorama respectively. For stereo panoramas captured by 360\degree\ cameras, we obtain their depths using the 360\degree\ depth estimation method proposed in \cite{feng2022depth360} to align the 3D scenes of the target and the source images. Then with the camera model introduced in \neil{Eq.~\ref{eqn:camera}}, for each pair of cameras $(C^i_L, C^i_R)$, we use \neil{their} view direction and the interval $\theta_i$ to obtain the affected pixels in the left and right views of the target panorama. Since the columns of an equirectangular image are naturally the pixels of different horizontal viewing angles, we just need to \neil{identify the affected} columns and project those spherical pixels to the 2D image plane of the synthesized image $I^r_i$ or $I^l_i$ to find the pixels to overwrite the original colors. The pixels' alpha values will be used to ensure that only valid pixels will be used and seamlessly blended. Finally, depending on whether the inserted region contains background pixels surrounding the object boundary, we optionally perform Poisson Blending \cite{perez2003poisson} on the synthesized regions in the left and interpolate the right view based on the geometric information.

We find that a larger number of segments always means better visual quality. Therefore, in our default settings, the viewing angle intervals are decided by the horizontal resolution of our target panoramic images and are normally set as the viewing angle represented by a single column. For example, for a panorama with a resolution of $3840 \times 1920$, \neil{we use $N=3840$ and} our interval is \neil{thus} $360^\circ/3840 = 0.09^\circ$. In our experiments, we found that the per-column synthesis will not introduce artifacts regarding the spatial continuity of the object's appearance, because the changes between neighbouring columns are subtle and neglectable.  
Finally, we \neil{can} compare the depth of the inserted object with that of the panorama to determine the occlusion relationship for better composition. 

\section{Experiments and Results}
In this section, we evaluate the key parts of our system, demonstrate our $360^\circ$ stereo image compositing results, and compare them with other stereo composition methods. We also conduct a user study to evaluate the depth perception quality of the generated results. 

\subsection{Implementation}
\label{Implementation}
The source object can be from any \neil{kind} of stereo images, such as images captured by 2D/360\degree\ stereo cameras or rendered stereo images, and single RGB-D images. \hk{Our system enables users to interactively mark their objects of interest and obtain alpha masks for them, and provides a ``one-off'' preview of the composition results on the target scene, allowing users to view the changes in real-time. Furthermore, the DDDN can be used to give users greater flexibility in editing the desired pose of the inserted object, such as orientation and scale. Once a dense depth map is obtained, it can be projected into 3D space and aligned with the target scene for per-view segmentation. The partitioned 3D points can then be projected onto the 2D image domain using perspective projection and densified using our proposed DDDN for each view. Finally, the equirectangular projection is applied to project the specific column on the predicted depth map and mask onto the target column of the 360 images, which we refer to as ``per-column composition''.}
\hkminor{Furthermore, for cases where there is a significant difference in pixel colors between the source object and the target region, Poisson Blending, as outlined in \cite{perez2003poisson}, may be optionally employed. While the left view can be generated using the aforementioned procedure, the right view can be synthesized by harnessing geometric information to obtain and interpolate corresponding pixel colors, thereby ensuring consistency in color and the underlying 3D geometry between both views.}

For consumer-grade stereo cameras for which the intrinsic parameters are available, we can directly use these to offer a better sense of the real-world size of the operated objects. For example, we used $f = 700$ and $B = 0.12$ for images captured by a ZED camera. Otherwise, we set some constant values for $f$ and $B$ for convenience and let the user adjust the object's size for the stereo data \cite{Xian_2020_CVPR} collected from the internet. When refining the final appearance, gradient-domain methods such as Poisson Editing are provided as an optional operation for users. 


\hk{When applying the per-column strategy on a single CPU core of an Intel Xeon W-2133 with an RTX3090 GPU, the average execution time of our Python implemented method is 45~seconds} for generating a composition result with a resolution of $3840\times 1920$ target 360\degree\ image ($N=3840$, 0.09\degree\ as the view interval and the object covers an FoV of 90\degree\  ), including object segmentation, depth estimation, point cloud processing, per-view projection and the final combine steps. 
\hk{The number of depth maps and masks that are produced is contingent on the number of columns (FoV) of the inserted object that is compositing on the equirectangular image. If a source object covers 90\degree\ of FoV in a $3840\times 1920$ 360\degree\ image, there will be a total of 960 depth maps generated for different camera views, in accordance with the original depth maps for each view.}

\hk{To minimize the computation time and ensure interactive performance, we implement \nd{an alternative ``key-column'' strategy that generates a smaller number of depth maps, each covering multiple columns of pixels. We found that one depth map per 11 columns of pixels works well without creating obvious discontinuities between one set of columns and the next. In the example above,} instead of generating 960 dense depth maps using the ``per-column'' strategy, \nd{the key-column strategy requires only 87 key columns for a 90-degree FoV coverage, each key column's depth map being used by the 5 columns of pixels to its left and the 5 to its right. The key-column strategy reduces} the execution time to 9 seconds, effectively reducing redundancy and computational costs, thereby making our approach interactive.}



\subsection{Component Evaluation}
\subsubsection{Real-world Generalization}
\hk{We use self-generated synthetic data to train our DDDN model for dense depth map generation, as there are no existing datasets that fully satisfied our requirements to the best of our knowledge. \nd{For example,} the 4D Light Field Dataset proposed by Honauer et. al \cite{honauer2017dataset}, \nd{while closely related to our problem, provides only} a single depth map for the default view location, despite having multiple views of the same object.}

\hk{It should be noted that, while synthetic data can be a useful starting point for the depth completion task, the trained model may not accurately capture the complexity and variability present in real-world data, as illustrated in Fig.~\ref{fig:domain_adaptation} (b). We incorporate \nd{additional} loss functions \nd{($L_b$ and $L_{bce}$)} to optimize the predicted depth map, accounting for the complexities and robustness required for real-world data. These losses ensure that the predicted depth map (masked) matches the default depth map after a corresponding reverse transformation, facilitating consistent completion across views while linking to the original input. By training with real-world data, we have observed significant \nd{improvement, as} shown in Fig.~\ref{fig:domain_adaptation} (c). }

\hk{We assess the effectiveness of our approach using standard metrics for depth map densification, including MAE, RMSE, SSIM, and PSNR, on a real-world dataset, as described in Section 4.3. Tab.~\ref{tab:rwgeneralisation} presents the quantitative results. The results demonstrate that our DDDN model, with the inclusion of real-world generalization, outperforms our original model in terms of these metrics. This validates the robustness and effectiveness of our proposed approach in real-world scenarios.}

\subsubsection{Comparing Rendering Methods}
\hk{We compare three different rendering schemes for 360\degree\ stereo image composition: per-column, key-column and one-off.}
\nd{The one-off scheme is the baseline against which we compare our new schemes (per-column and key-column). The one-off scheme is a 2D stereo composition method that takes the stereo contents as input and applies one-off object insertion operations on the left and right views of the target image. This produces incorrect stereo disparities over large parts of the 360\degree\ image. In contrast,} \hk{our new schemes are able to generate correct depth perception.} A basic requirement for a stereo image pair is that there should be only horizontal disparity for each corresponding pixel pair when the viewer is looking at that point, to fit the layout of human eyes. 
\hk{However, the stereo 360\degree\ images created using the one-off method \nd{provide roughly accurate disparity results only} for the area directly in front of the capture cameras. Behind the cameras, the perspectives generated \nd{are} reversed, and \nd{to the left and right} of the cameras, there \nd{is} vertical disparity. This is because the two cameras are placed at fixed positions, spaced apart by the average distance between a person's eyes. \nd{Looking directly ahead, their separation is correct, left--right. But at 90\degree\ to that direction, the relative positioning of the \nd{two cameras} is not left--right but instead front--back, resulting in a scale change rather than a stereo disparity: this is incorrect, does not provide correct stereo perception, and can result in eye strain, \hkminor{such as Fig.~\ref{fig:projection_comparison}}. Indeed, the} one-off method produces displacements in all directions. The direction \nd{and magnitude} of displacement depends on the object's \nd{distance from and angle to the pair of cameras. To ascertain the scope of this problem and whether our schemes resolve it, we} conducted a quantitative analysis of the disparities generated by \nd{the three schemes,} comparing them against the disparities present in the ground truth (GT) stereo 360\degree\ image pair generated by Unity3D.}

\hk{We use Unity3D to generate a chessboard that covers 119.25\degree\ FoV in the final stereoscopic results with the correct depth adaption across all the covered regions. Using such synthetic data, we are able to avoid possible 3D reconstruction errors when working on real-world data, so that the comparison can focus only on the generated disparities.} For the one-off \nd{scheme}, we use only the initial left and right camera positions described in Sec.~\ref{sec:ViewSeg} and render two 360\degree\ images as the stereo pair. \nd{For our per-column and key-column schemes,} we segment the 3D scene as in Sec.~\ref{sec:ViewSeg} according to the view angle range covered by \hk{a number of columns} of the panorama and render the pixels within the view range using the pair of cameras looking at that direction. Then the per-column or key-column results are combined together to form the final stereo panorama with depth-adapted left and right content.

\hk{Tab.~\ref{tab:renderingevaluation} presents a quantitative analysis of the disparities generated by the per-column, key-column (11~columns), and one-off \nd{schemes}. We compare the average Euclidean distance between disparity vectors against the GT stereo 360\degree\ image pair produced by Unity3D at the centered different positions \nd{($\theta = 0^\circ$, $\phi = -70^\circ, -35^\circ, 0^\circ, 35^\circ, 70^\circ$)}. Our \nd{schemes} achieves significantly lower Euclidean distance differences than the one-off \nd{scheme} on the disparity vector between the left and right views of the covered region, which shows the generalization ability of our method when compositing to different regions on a sphere. It is worth noting that the key-column \nd{scheme} \nd{generates} comparable evaluation results to the per-column approach, while substantially decreasing redundancy and computation time in comparison.}

\begin{table}
\color{black}
\centering
\normalsize
\begin{tabular}{cccc}
\hline
\multicolumn{1}{c|}{Method} & \multicolumn{1}{c|}{Per-column} & \multicolumn{1}{c|}{Key-column} & \multicolumn{1}{c}{One-off} \\ \hline
\multicolumn{1}{c|}{Disparity Difference} & \multicolumn{1}{c|}{\textbf{0.6544}} & \multicolumn{1}{c|}{\textbf{0.7733}} & \multicolumn{1}{c}{\textbf{3.0327}} \\ \hline
\end{tabular}
\caption{\hk{Quantitative results on corner points of a chess board with different rendering methods: per-column, key-column (11 columns), and one-off.}}
\label{tab:renderingevaluation}
\end{table}


\subsubsection{Deep Depth Densification}
\zfl{Our deep model for depth densification is built to generate a dense depth map from a sparse projected point cloud. To demonstrate the necessity of the dedicated deep network, we compare it with two possible alternatives: depth map completion methods and point cloud completion methods. Most depth map completion methods are designed to improve the depth map quality of a given RGB-D image. They need a complete RGB image to guide the depth completion, which is not available in our task because the dense RGB image for the desired pose is also missing. Thus, we choose to compare with the method of \cite{ku2018depthCompletion} because it can perform depth completion based on only an input sparse depth map. }   

\zfl{Some results of the depth completion method proposed in \cite{ku2018depthCompletion} are shown in the third row of Fig.~\ref{fig:compareDepth}. Due to the limitations of their morphological operations when filling the missing pixels, their method fails to maintain geometric details, e.g., the trees' edges in the first example in Fig.~\ref{fig:compareDepth}. Their method can also easily propagate incorrect depth values to its neighbours, causing undesirable depth effects like in the boundary regions of the second and fifth examples. Finally, their method may generate incorrect object shapes as in the third and fourth examples, because their method cannot predict which positions have valid pixels of the foreground object. Our deep depth densification method overcomes the above issues and produces higher-quality depth maps. It should be also noted that our method is 6 times faster than the method of \cite{ku2018depthCompletion}.}

\zfl{In the comparison between point cloud completion and our deep depth densification, we feed the transformed sparse point cloud to one of the state-of-the-art deep point cloud completion methods, PointTr~\cite{yu2021pointr} and project it to generate the depth map for the target view. We found that the point cloud completion method focuses more on the global 3D structure and the integrity of the model and fails to generate sufficient depth details for a specific view. The mandatory sampling step of their method is also a reason for the failure of the dense depth map generation. It cannot guarantee that all the geometric details of the depth map are preserved after the sampling step. Some typical examples are shown in Fig.~\ref{fig:compareDepth}, where our method generates depth maps with significantly better visual quality than PointTr. 
}

\zfl{We also consider NeRF-based approaches to directly learn to generate novel views. Unlike most NeRF-based approaches that need a number of input views to reconstruct a neural radiance field, PixelNeRF~\cite{yu2021pixelnerf} and Depth-Supervised NeRf~\cite{deng2022Nerf}  only need one or few input images to synthesize new images of novel perspectives. However, } \hk{PixelNeRF and DSNeRF algorithms face challenges in generating high-quality textures from stereo pairs, as the short stereo baseline typically leads to limited depth constraints estimated through structure-from-motion. This limitation results in unsatisfactory visual quality for composition tasks. Moreover, the algorithms often recover incomplete point clouds, typically only capturing the surface facing the camera, which further limits the ability to generate clear and sharp textures.}

\subsection{Results}
\begin{figure}[t]
    \centering
    \includegraphics[width=0.48\textwidth]{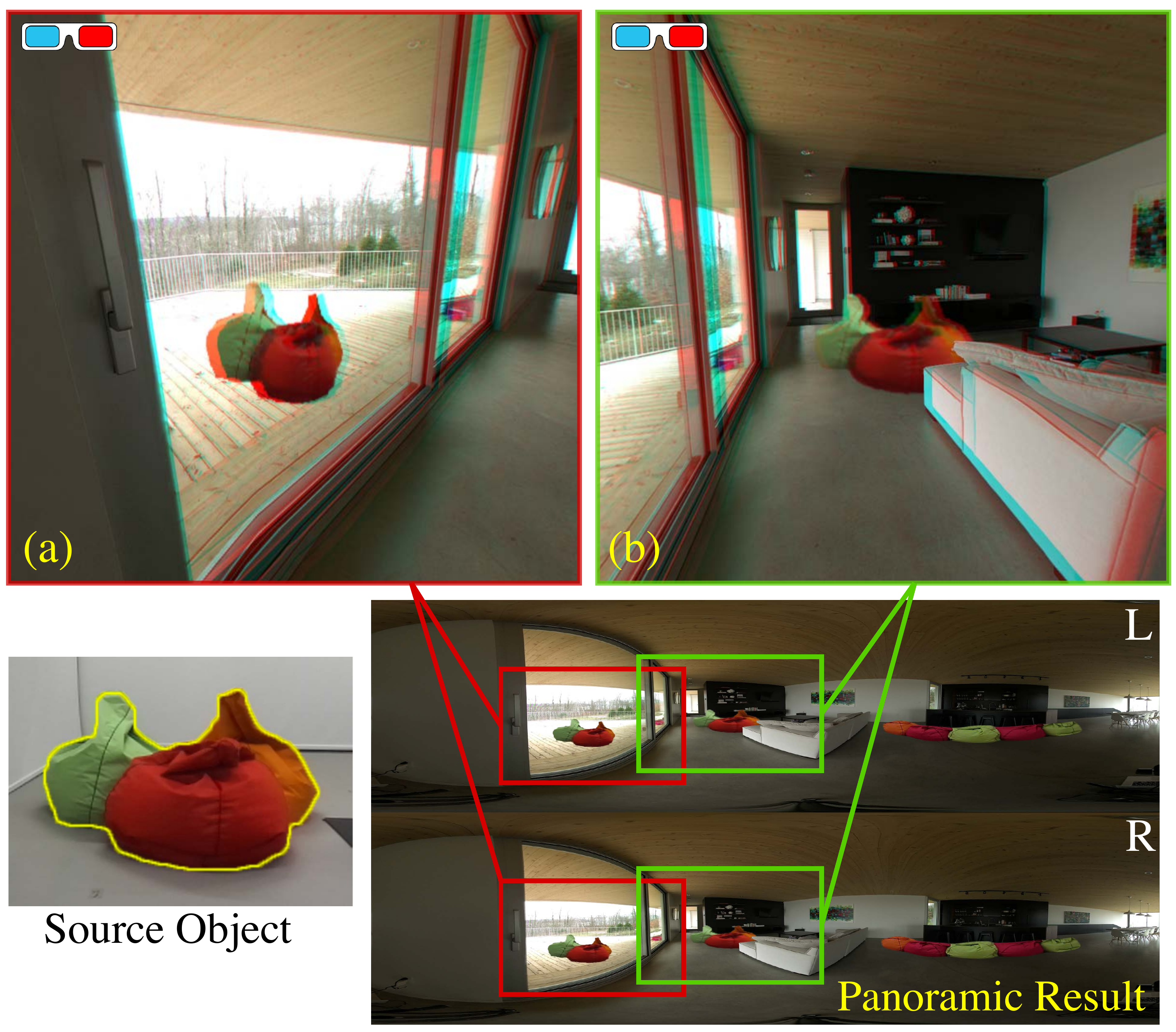}
    \caption{\hkminor{Composition results by the proposed system. (a) Foreground objects are rotated about the y-axis, and the red beanbag blocks the corresponding region of the orange one. (b) Composited foreground objects are occluded by the existing objects based on their depths.}}
    \label{fig:foregrounds_editing}
\end{figure}

\begin{figure*}[t]
  \includegraphics[width =1\linewidth]{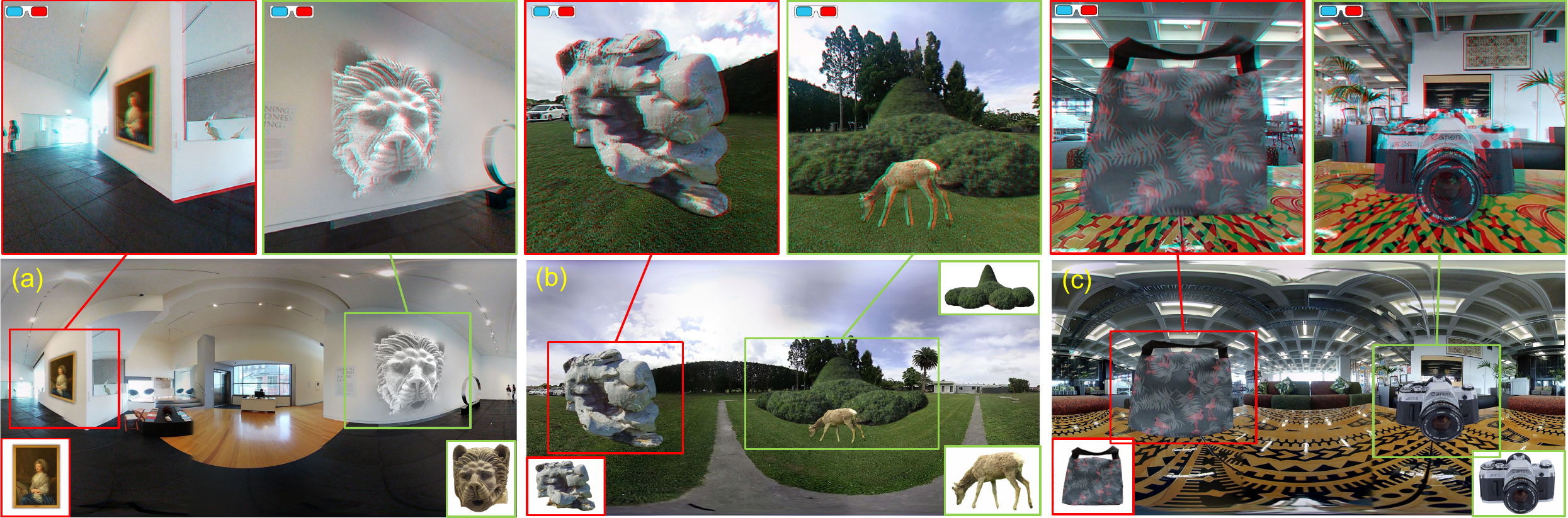}
  \includegraphics[width =1\linewidth]{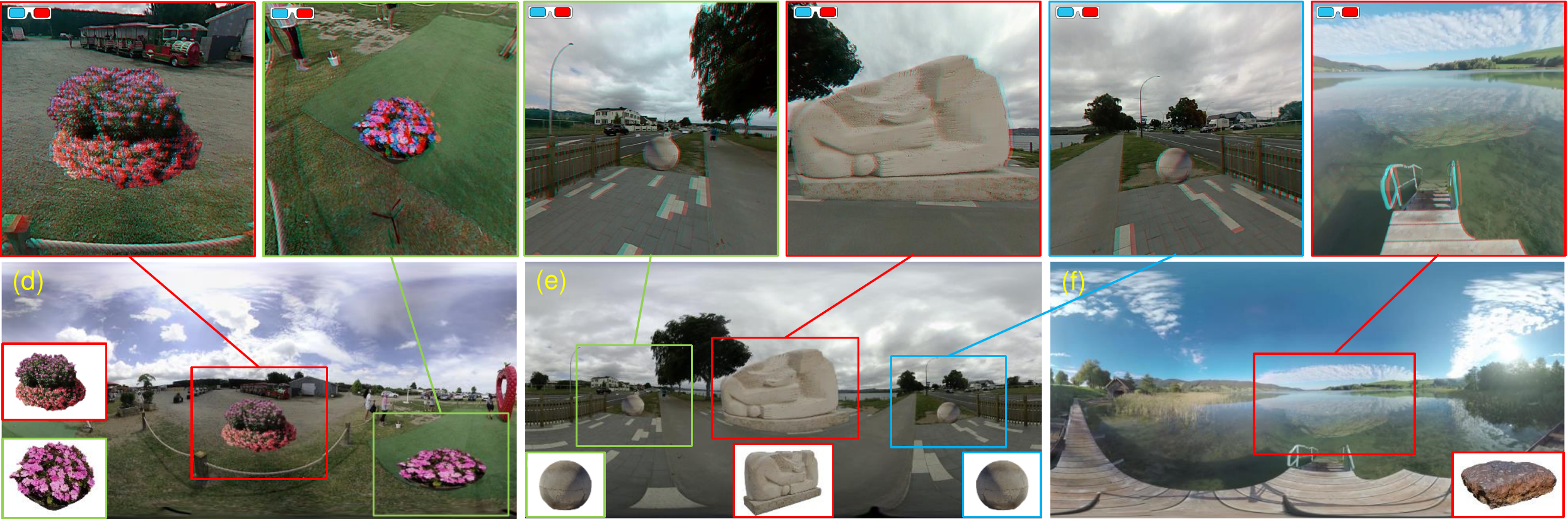}

    \caption{\zfl{Results of the proposed method. We show the anaglyph images of the zoom-in windows of the composited objects and only include the left view of the stereo panoramic composition results.}} 
  \label{fig:finalRes}
\end{figure*}
\begin{figure*}[t]
  \includegraphics[width =1.00\linewidth]{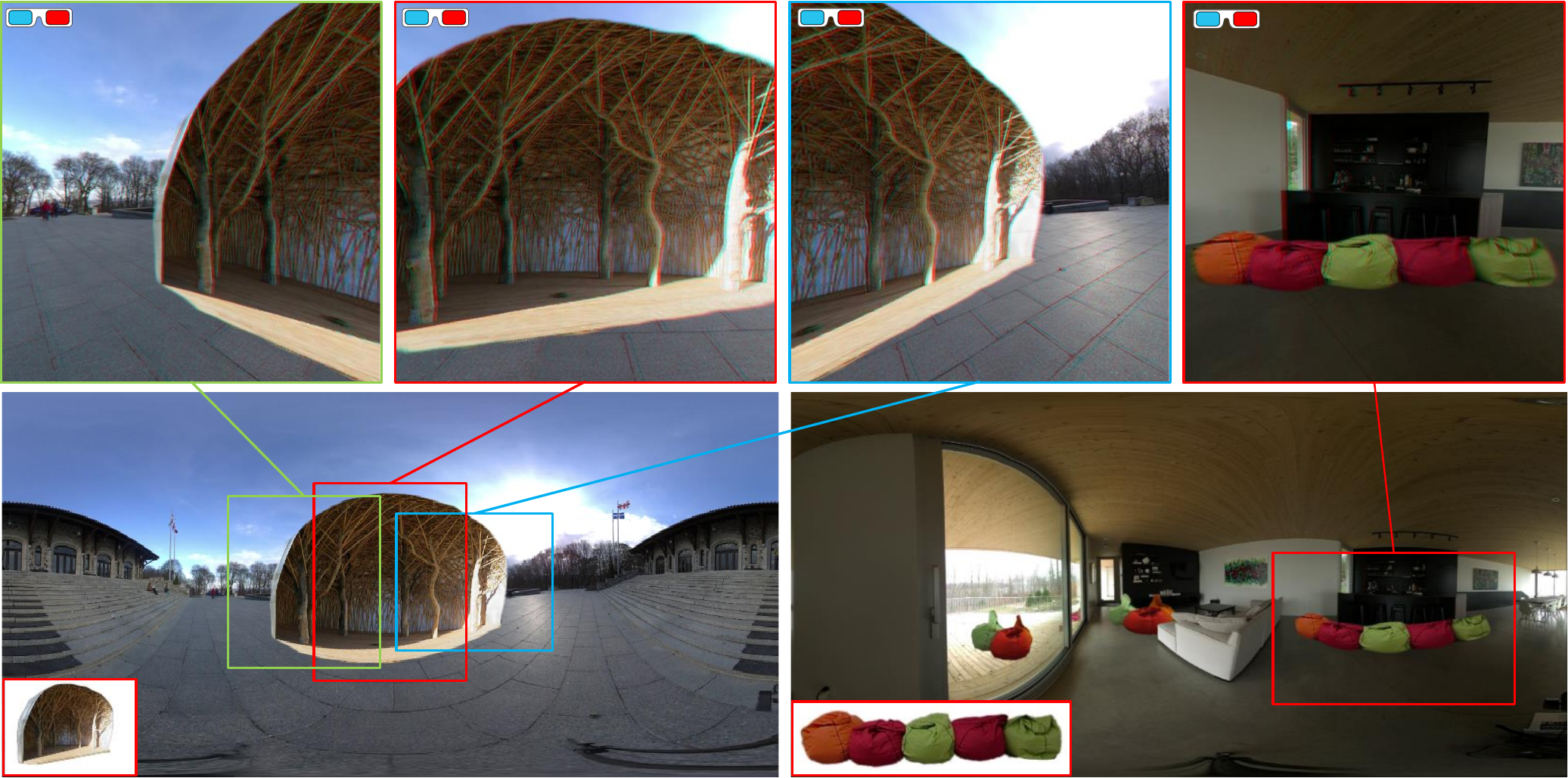}
  \vspace{-0.2in}
  \caption{\hkminor{Composition results with long objects inserted. Our method can allow arbitrary sizes of insertions covered in the target scene.}}
  \label{fig:largeFoVRes}
\end{figure*}

\zfl{Some composition results of our approach are presented in Fig. \ref{fig:finalRes}. For each example, the stereo panoramic scenes are shown at the bottom using the equirectangular projection of its left view. The inset windows demonstrate the segmented source objects and the zoom-in windows visualize the composited stereo objects using anaglyph images. We also include all the stereo 360\degree\  results in our supplementary materials, which can be viewed with VR headsets to achieve a better depth perception. From the results where we change the objects' orientation, size and position, we can see that our method is suitable for processing panoramic images as it is able to recover and manipulate the 3D geometry information to guide the pixel generation. We naturally avoid the computation for adapting to equirectangular distortions when pasting the object into an arbitrary position. We also generate correct panoramic disparities using this 3D-guided approach. Fig. \ref{fig:foregrounds_editing} shows a result where we achieve natural composition results by considering the occlusion relationship with the backgrounds.
The occlusions between the composited objects and the original object show the depth consistency achieved by our method.}
\hk{In Fig. \ref{fig:finalRes} (a) and (f), we apply Poisson Blending to one view to achieve natural color matching with the target scene, and then leverage geometric information \hkminor{to transfer and interpolate corresponding pixel colors from one view to the other through the image warping (point cloud-based) technique}
, thus ensuring color and underlying 3D geometry consistency between both views.}
\hk{In addition, our method has no restrictions on the size of the source object or the FoV \nd{covered} in the target scene, which is particularly suitable for 360\degree\ image editing tasks. One reason is our proposed work manipulates the point cloud in the 3D space and then projects it to the target panoramic images. The other reason is that the target panoramic (equirectangular) image already covers \nd{$360^\circ\times180^\circ$} of a scene, so the target 360\degree\ images will always be able to \nd{incorporate} any size of composited object. Fig. \ref{fig:largeFoVRes} shows two such examples, where our approach achieves correct disparities in the left and right end of the composited objects.}



\begin{figure}
\centering
    \includegraphics[width=\linewidth]{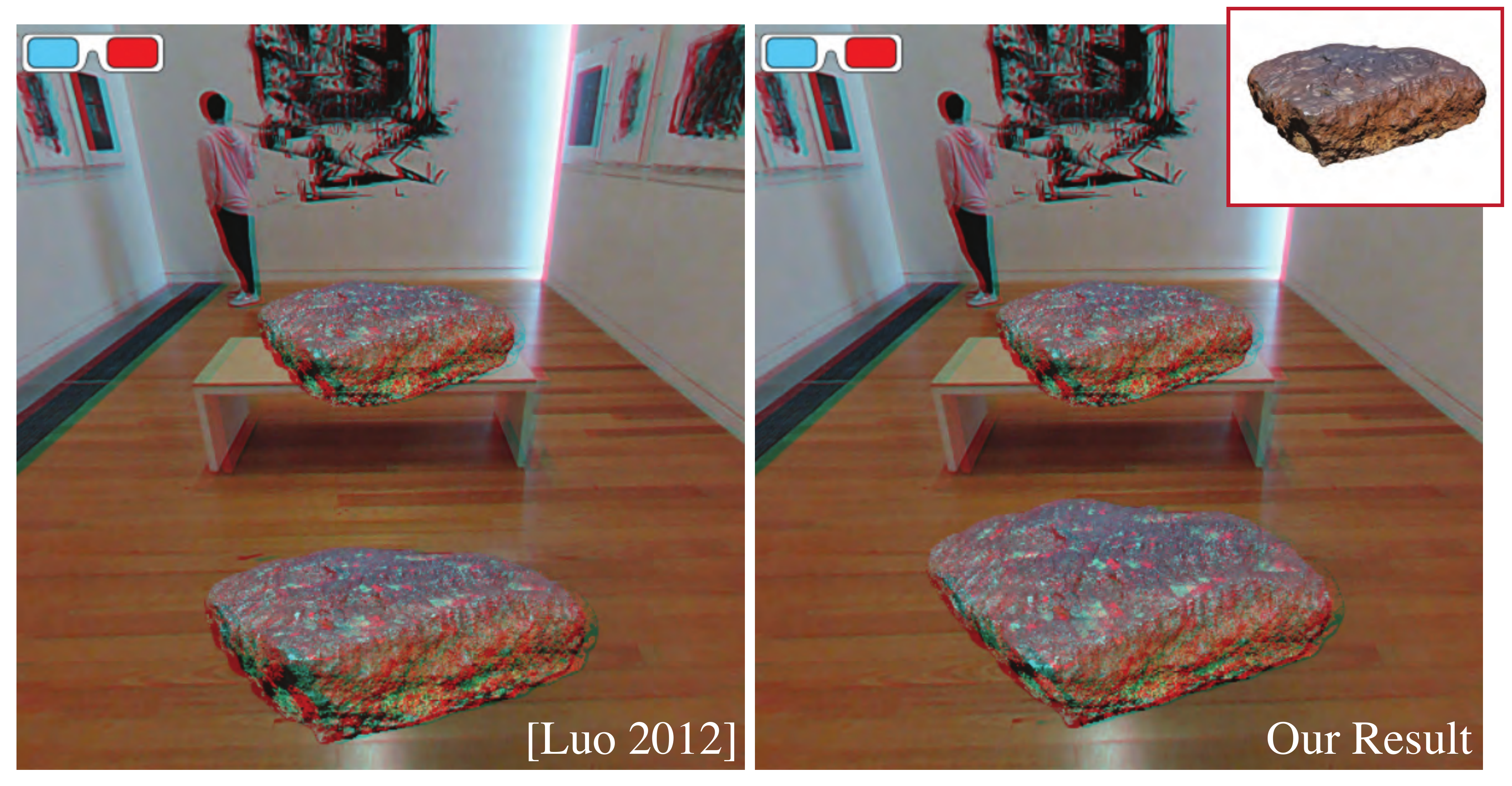}
  \caption{Comparing our method with the 2D stereoscopic image composition method described in \cite{luo2012perspective}. We inserted two rocks into the target scene at different depths and heights. Note that our method can adaptively adjust the 3D pose of the inserted object for a more natural result.}
  \label{fig:compareLuo}
\end{figure}

\begin{figure}[t]
    \centering
    \includegraphics[width=0.48\textwidth]{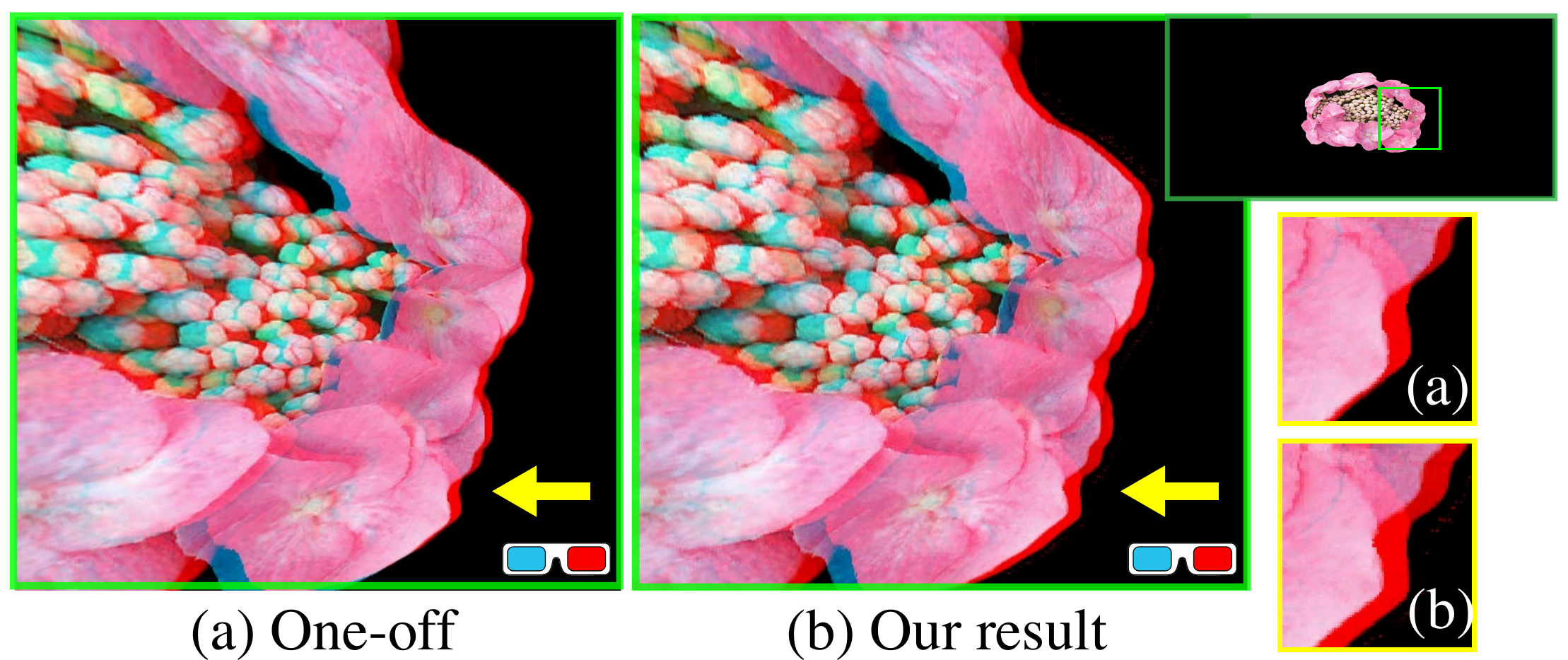}
    \caption{Rectilinear views of synthesized stereo objects. It can be seen in the anaglyph images that the one-off method can not guarantee the horizontal disparities required for correct depth perception.} 
    \label{fig:projection_comparison}
\end{figure}


\subsection{Comparisons}
Directly treating equirectangular images as normal 2D images when applying the image cloning operation cannot generate correct equirectangular distortion, which is important for maintaining the inserted object's shape when viewing it in a headset or a 360\degree\ image player. 
Considering the incorrect equirectangular distortion will also lead to problematic disparities when viewing the region of interest of the result, we therefore do not make further comparisons with the 2D planar image composition methods proposed for monocular 2D images. 

For objects that only cover a narrow FoV, an alternative method to insert stereo objects to the target panoramic scene is \neil{to project} the relevant part of the panorama to a 2D image plane and \neil{to then insert} the object \neil{into} the planar stereo image before projecting back. The composition methods of Luo et al.~\cite{luo2012perspective} and Tong et al.~\cite{tong2012stereopasting} are proposed for image cloning and composition in stereo 2D images. The latter method needs a substantial amount of user interaction. Therefore, we choose to compare with the method of Luo et al. that relies on mesh-based deformation to demonstrate the effectiveness of our stereo content manipulation and generation method. In Fig.~\ref{fig:compareLuo}, we show their composited stereo images and the 2D result generated using our deep depth densification and view-dependent content generation. Due to the limited capability of the mesh-based deformation method on the perspective changes of the object, our method can produce more realistic results when the desired orientation and relative position of the target object \neil{is notably different from the original capture}. 

\zfl{Fig.~\ref{fig:projection_comparison} shows examples of both the one-off method and our method. Viewing the anaglyph stereo images, it is clear that our method produces only desirable horizontal disparities, while the one-off method produces incorrect results that include undesirable vertical disparities. We evaluate the perceived visual quality of the two methods in our user study.}

\subsection{\nd{Alternatives and additions to the algorithm}}\label{sec:morediscussion}
\nd{Pixel color inpainting is a useful technique in composition. In our case, it might lead to higher quality results, but it raises additional research questions. We have not applied any pixel inpainting technique onto our incomplete inserted objects because our focus is primarily on maintaining the composited object's disparity consistency across all view directions in the 360\degree\ images. If one did wish to apply inpainting, then a substantial unsolved issue is how to handle the inconsistency in the occluded regions for the left and right eyes. This inconsistency makes it crucial to prioritize creating a consistent color pattern in each view during any inpainting process. Any pattern mismatch between views could negatively impact the correctness of the disparities. Rather than use inpainting, we addressed the missing region color issue through our \hkminor{point cloud-based} image warping technique that uses the predicted depth map to retrieve pixel color from corresponding positions.}

\hk{\nd{We recognise that} using the low-quality estimated depth map and alpha mask as input \nd{can} result in imperfect final composited results. For example, if the depth estimation technique fails to accurately estimate the depth of the object, the foreground object might appear distorted or disconnected from its surroundings especially when its orientation is changed. Similarly, if the alpha matting process is not accurate, the edges of the foreground object might appear jagged or rough, resulting in an unnatural-looking composite image. To overcome these issues, we adopted one of the state-of-the-art depth estimation techniques \cite{li2020revisiting} which delivers highly precise depth estimates for objects. \nd{Where this proves to be insufficiently accurate,} we can encourage users to use some incorporated user interaction masking techniques to further improve the accuracy of the input mask.}

\hk{\nd{In addition, we noted that low-quality depth maps can lead to floating points around the new shape after the object's orientation is changed. Our} depth densification network denoises these floating points, which provides more tolerance for low-quality depth maps. We also include preprocessing techniques on the depth map that reduce noise points in the input. \nd{These additions to the algorithm} enhance input accuracy \nd{so that} our proposed system produces more realistic and satisfactory outcomes. }

\nd{Finally, an alternative approach would be to use 3D point clouds rather than depth maps. However, we see several challenges with this approach.} \hk{For instance, if we segment views based on the density of the point cloud, it could disrupt the disparity coherence of the inserted object in the final ODS image because the appropriate disparity for any given part of the inserted object should be determined by the user's viewing direction, rather than by the density of the point cloud itself.}

\nd{The task of completing the depth map itself could be addressed by a point cloud upsampling technique. However, such upsampling algorithms primarily aim to enhance the global 3D structure and integrity of the inserted object. As an example, consider the method of Liu et al. \cite{liu2022pufa}. It takes a point cloud of 256, 1024, or 4096 points, and densifies this by a factor of four, so a maximum of 16384 points. On the other hand, our depth densification network has a $512\times512$ depth map as both input and output. This is a considerably higher number of points than in the point cloud upsampling method.} \hk{As a result, a complete single view of the inserted object is sufficient for us to reconstruct, and our approach can provide much finer details at the local level. }

\begin{figure}
\centering
 \includegraphics[width =1\linewidth]{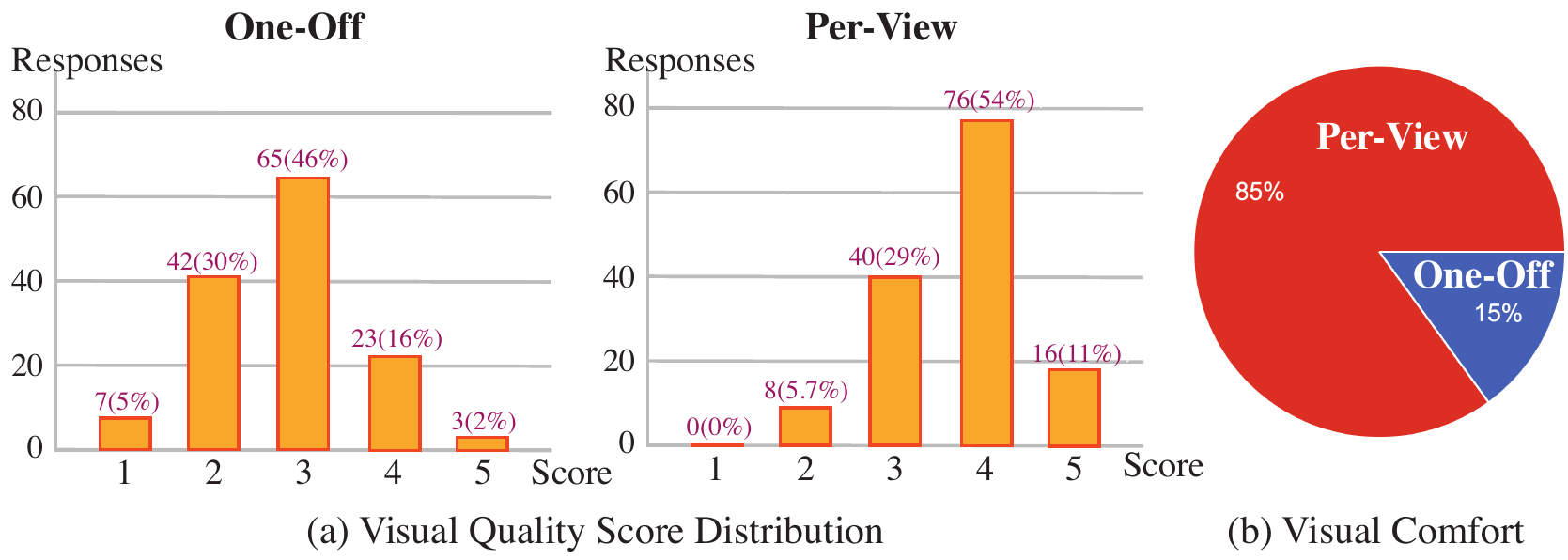}
 \caption{The user study results. (a) The visual quality score distributions of the two methods (on scale of 1-5). (b) The visual comfort preference.}
  \label{fig:userstudy}
\end{figure}

\subsection{User Study}\label{sec:userstudy}
\zfl{We conduct a user study to validate our method subjectively. We used ten panoramic images of different foreground objects. We generated two results for each scene: one result where we directly project all 3D points to the left and right equirectangular image to guide the pixel generation with a fixed pair of cameras (one-off) and the other result using our per-view projection, where we choose the FoV of each column of a $3840\times 2160$ panoramic image as the interval of our view-segmentation, i.e., 0.09\degree. There were 14 participants (six males, and eight females, aged from 25 to 52). After a short training session using two stereo scenes, the participants were asked to watch the ten groups of two stereo panoramas in Oculus Quest 2, and assess the two panoramas based on the visual quality by giving a score from 1-bad to 5-good. They were also asked to mark three positions where they found the visual qualities were most different between the two shown panoramas. Finally, they were asked to choose one of the two compared panoramas that had better visual comfort. We presented the foreground objects in front of a black background to avoid any confounding factors from a textured or image-based background.} 

\zfl{Fig.~\ref{fig:userstudy} and Tab.~\ref{tab:userstudy} report the user study results. The per-view result achieves a much higher mean score than the one-off results which may be attributable to no ghost artefacts being perceived by participants, especially at the boundary or the regions with sharp and clear structures. This is consistent with our intuition that using a large number of segments with small intercepts can benefit the capture of foregrounds by rendering all parts of the foreground from appropriate eye positions. We performed a paired-sample t-test between the scores of the results of the one-off method and our approach. As shown in Tab.~\ref{tab:userstudy}, the result indicates that the visual quality of our method is significantly better than the one-off method at a significant level $\alpha = 0.05$. Most of the participants reported that the regions far away from the composited object's center have noticeable quality differences. Depending on the textures and colors, the participants might have a different level of sensibility to such a difference, which causes a minor variance in the reported positions. In terms of visual comfort preference, the preferred method is our per-view approach in 85\% of the responses. For more information on the details and results of our user study please refer to our supplementary document. The above user experiments have been approved by the Human Ethics Committee of Victoria University of Wellington (ID: 0000025362).
}

\begin{table}
\centering
\normalsize
\begin{tabular}{cccc}
\hline
\multicolumn{1}{c|}{Score}          & \multicolumn{1}{c|}{Mean} & \multicolumn{1}{c|}{Std} & \multicolumn{1}{c}{P-Value}                              \\ \hline
\multicolumn{1}{c|}{One-off}        & \multicolumn{1}{c|}{2.807}  & \multicolumn{1}{c|}{0.847} & \multicolumn{1}{c}{\multirow{2}{*}{$5.607\text{e-}20$}} \\ 
\multicolumn{1}{c|}{Per-column} & \multicolumn{1}{c|}{3.714}  & \multicolumn{1}{c|}{0.742} & \multicolumn{1}{c}{}                    \\ \hline            
\end{tabular}
\caption{Statistics of the user study results. The paired-sample t-tests are performed between the scores of the two methods. }
\label{tab:userstudy}
\end{table}

\ \ 

\noindent\textbf{Limitations and Future Work}  
\zfl{
The proposed approach has three limitations. First, if the depth estimation method fails to predict an accurate depth map of the source stereo object, our image generation method based on depth maps might not be able to produce satisfactory results when the user wants to change the object's 3D pose due to the incorrect 3D-to-2D projection. \hk{Second, our approach does not estimate the illumination of the target scene and the composited object. In future work, we will reconstruct the lighting information and the 3D geometry of the target scene to illuminate the object for better consistency. } Finally, our current system relies on the user's input to decide the object's 3D pose and position. More advanced pose adjustment methods can be potentially employed to create more realistic results.}

\section{Conclusion}
The goal of this paper is to address stereo 360\degree\ image composition with desired poses and positions of the inserted object, especially when the user composites an object with desired pose and scale that covers a large FoV into an ODS image. The goal has been achieved by developing a novel composition algorithm that keeps the basic 3D geometry of the composited object, while also achieving a high-quality depth perception for an arbitrary view in the panoramic scene. Particularly, a per-view projection method can make the composited content adapt to different view directions. The results show that the composited foregrounds can keep geometry information when the perspective, position or size of the object change. The user study demonstrates our method achieves the highest quality of depth perception when we make the per-view projection method with a fine segmentation. 

\ifCLASSOPTIONcompsoc
  \section*{Acknowledgments}
\else
  \section*{Acknowledgment}
\fi

This work was supported by Marsden Fund Council managed by Royal Society of New Zealand (No. MFP-20-VUW-180). Thanks to the Te Pataka Toi Adam Art Gallery for sharing their space with us for taking photos for Fig. 9(a) and Fig. 11.


\bibliographystyle{abbrv-doi}

\bibliography{main_TVCG}

\begin{IEEEbiography}[{\includegraphics[width=1in,height=1.25in,clip,keepaspectratio]{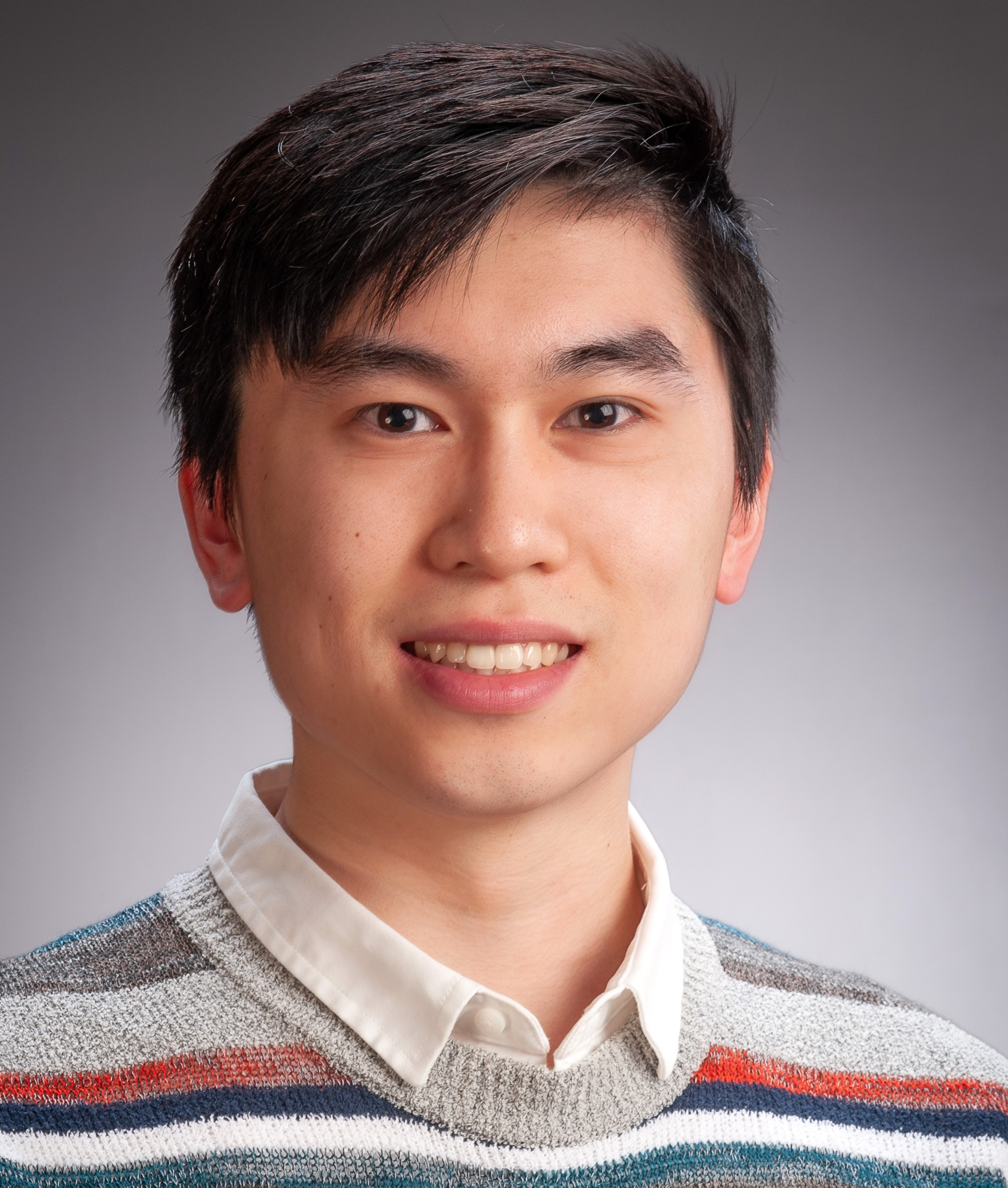}}]{Kun Huang} is currently a PhD candidate with Victoria University of Wellington, New Zealand. He received a Bachelor's and M.S. degree from Victoria University of Wellington in 2017 and 2021, respectively. His research interests include 360 image and video editing, virtual reality and mixed reality. He is a member of IEEE. He is a student branch chair of the IEEE New Zealand Central Section. 
\end{IEEEbiography}

\begin{IEEEbiography}[{\includegraphics[width=1in,height=1.25in,clip,keepaspectratio]{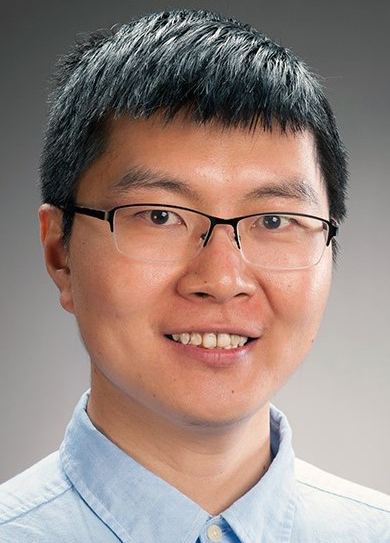}}]{Fang-Lue Zhang} is currently a senior lecturer with Victoria University of Wellington, New Zealand. He received a Bachelor's degree from Zhejiang University, Hangzhou, China, in 2009, and a Doctoral degree from Tsinghua University, Beijing, China, in 2015. His research interests include image and video editing, mixed reality, and image-based graphics. He is a member of IEEE and ACM. He received Victoria Early Career Research Excellence Award in 2019. He is on the editorial board of Computer \& Graphics. He is a committee member of the IEEE New Zealand Central Section. 
\end{IEEEbiography}

\begin{IEEEbiography}[{\includegraphics[width=1in,height=1.25in,clip,keepaspectratio]{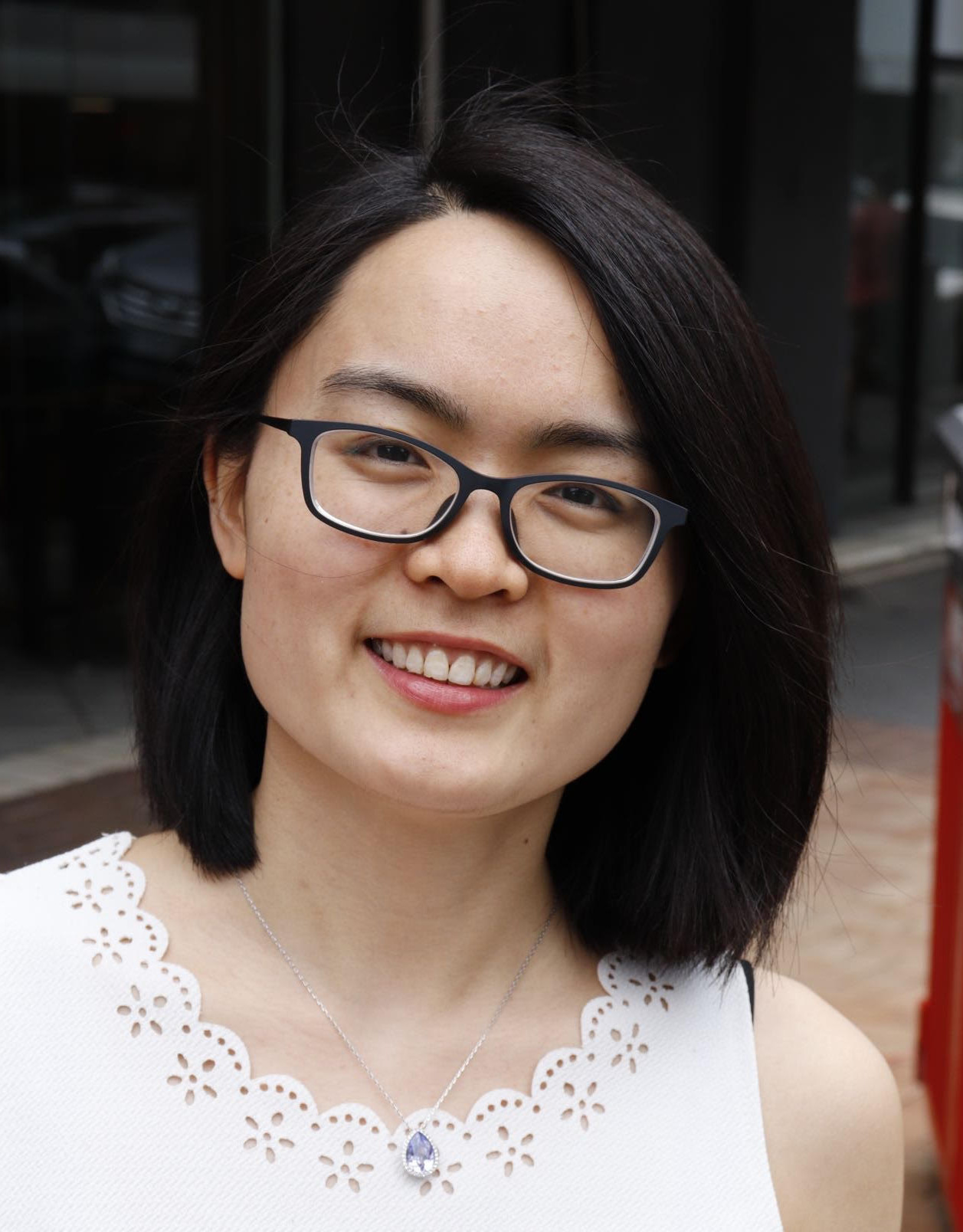}}]{Junhong Zhao} is currently a Research Fellow with the School of Engineering and Computer Science of Victoria University Of Wellington. She completed her doctoral degree in 2015 at the Institute of Electronics of the Chinese Academy of Sciences. She worked at the Institute of Information Engineering of the Chinese Academy of Sciences as an Assistant Researcher from 2015 to 2017. From 2018 to 2022, she was working with the Computational Media Innovation Centre (CMIC) at Victoria University of Wellington as a postdoctoral research fellow. Her research interests include machine learning, image processing and computer vision. 
\end{IEEEbiography}

\begin{IEEEbiography}[{\includegraphics[width=1in,height=1.25in,clip,keepaspectratio]{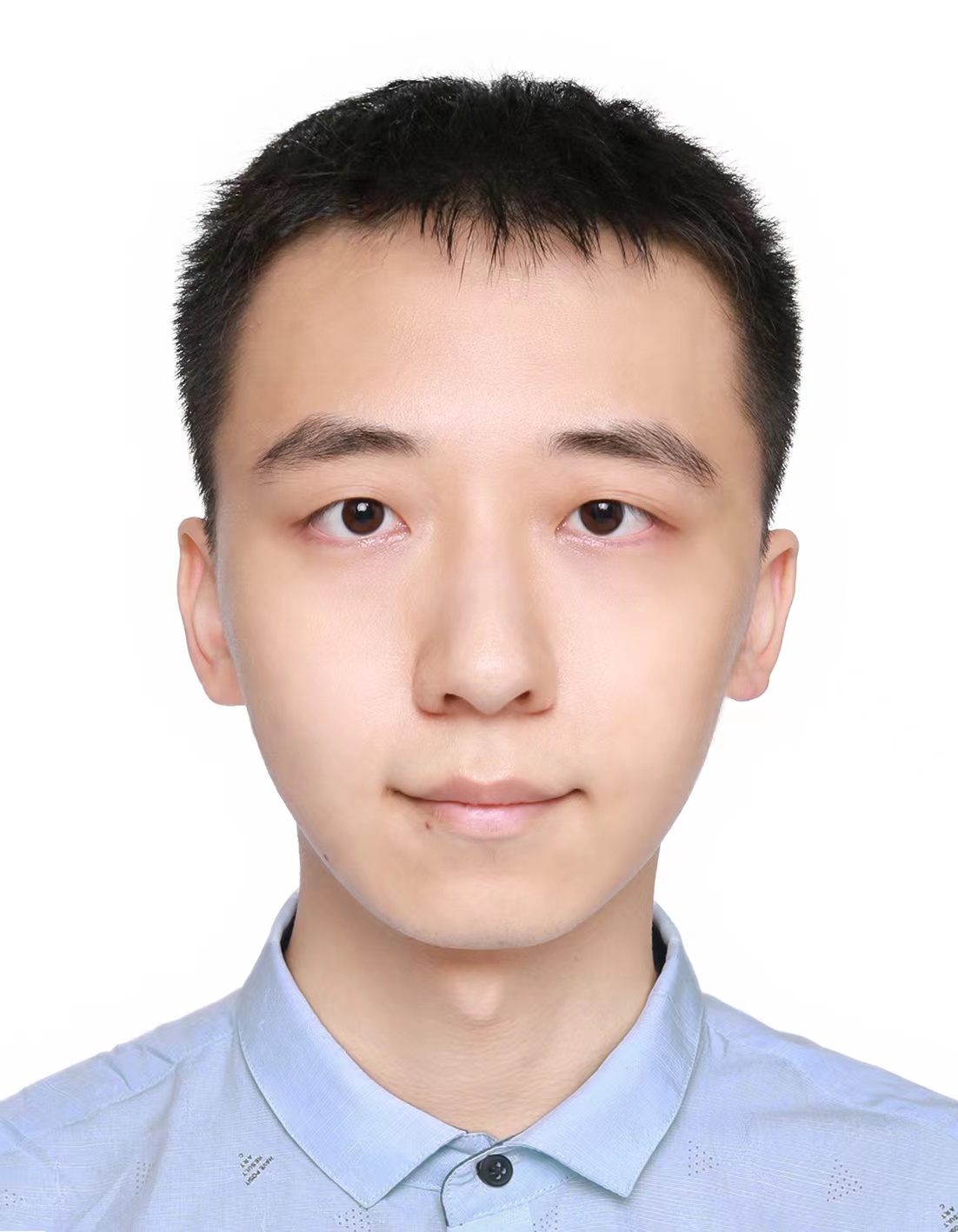}}]{Yiheng Li} received the M.S. degree in Computer Graphics in 2022 from Victoria University of Wellington, New Zealand. He is currently working at Sony China Software Center. His research interests include synthetic dataset generation, high-performance computing and panoramic image processing.
\end{IEEEbiography}

\begin{IEEEbiography}[{\includegraphics[width=1in,height=1.25in,clip,keepaspectratio]{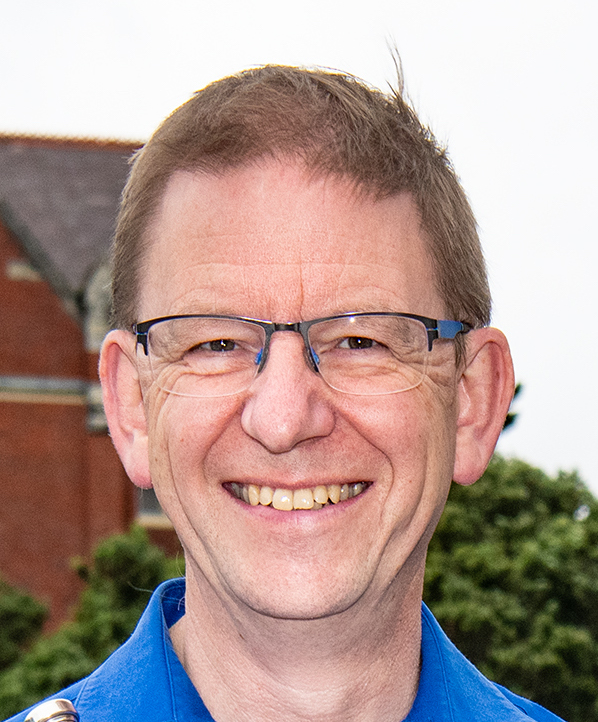}}]{Neil Dodgson} is Professor of Computer Graphics and Dean of Graduate Research at Victoria University of Wellington. His PhD is in image processing, from the University of Cambridge. He spent 25 years at Cambridge, becoming full professor in 2010. He moved to Wellington in 2016 to lead the computer graphics group there. His research is in 3D TV, subdivision surfaces, imaging, and aesthetics. He is a Chartered Engineer and a Fellow of Engineering New Zealand and of the Institution of Engineering and Technology (IET) and the Institute for Mathematics and its Applications (IMA) in the UK.
\end{IEEEbiography}
\end{document}